\documentclass[preprints,article,accept,moreauthors,pdftex]{Definitions/mdpi}
\usepackage{graphicx}
\usepackage{natbib}
\usepackage[normalem]{ulem}
\firstpage{1}
\pubvolume{1}
\issuenum{1}
\articlenumber{0}
\pubyear{2022}
\copyrightyear{2022}
\datereceived{}
\dateaccepted{}
\datepublished{}
\hreflink{https://doi.org/10.3390/universe8100521} 

\Title{Outliers in spectral time lag selected gamma-ray bursts}

\TitleCitation{Outliers in GRBs}

\Author{Fei-Fei Wang $^{1}$, Yuan-Chuan Zou $^{2}$\orcidA{}}

\address{%
$^{1}$ \quad School of Mathematics and Physics
Qingdao University of Science and Technology
Qingdao 266061, China; wangfeifei@qust.edu.cn\\
$^{2}$ \quad School of Physics
Huazhong University of Science and Technology
Wuhan 430074, China; zouyc@hust.edu.cn}

\corres{Correspondence: zouyc@hust.edu.cn}

\abstract{It is possible that the astrophysical {samples} are polluted by some outliers, which might belong to a different sub-class. By removing the outliers, the underline statistical feature may be revealed. {A more reliable correlation can be used as a standard candle relation for the cosmological study.} We present outlier searching for gamma-ray bursts with Partitioning Around Medoids (PAM) method. In this work, we choose three parameters from the sample, while all of them having rest-frame spectral time lag ($\tau_{\rm lag,i}$). In most cases, the outliers are GRBs 980425B and 030528A. Linear regression is carried out for the sample without the outliers. Some of them have passed hypothesis testing, while others have not. However, even for the passed sample, the correlation is not very significant. More parameter combinations should be considered in the future work. }


\begin{document}

\section{Introduction}

Gamma-ray bursts (GRBs) are astronomical phenomena detected at high energies. Classification and correlation seeking may reveal the underlying physics. GRBs {can be classified as} long GRBs (LGRBs) and short GRBs (SGRBs). SGRBs are likely associated with compact binary coalescence events involving at least one neutron star \citep{Abbott2017, Goldstein2017}. Massive stellar collapse is believed to generate the LGRBs \citep{Woosley1993, Woosley2006, Blanchard2016}. However, there is no clear criterion for the classification. Some GRBs could not be decisively classified in either class. On the other hand, it is still possible that GRBs can be classified into more groups, such as intermediate class, giant flare from soft gamma-ray repeaters. The outlier detection belongs to classification analysis. The outliers may belong to an individual group.

Cluster analysis techniques are various, such as k-means, K-medoids, hierarchical clustering, neural network clustering, kernel principal component analysis \citep{Balastegui2001, Chattopadhyay2017, Modak2017}. \citet{Modak2021} conducted the fuzzy clustering on the GRBs from the final Burst and Transient Source Experiment (BATSE) catalog, and confirmed three groups. Partitioning Around Medoids (PAM) is the most prominent method of K-medoids. We use the euclidean distance as the similarity measurement. The outliers are far away from the other data. We analyze the outliers, and seek the underlying linear correlation between different properties.

The data is based on the collection of \citet{Wang2020}. We collect a full sample including prompt emission, afterglow and host galaxy properties. We choose arbitrary three parameters based on the rest-frame spectral time lag $\tau_{\rm lag,i}$ to find the outliers, i.e., each combination of the physical parameters contains $\tau_{\rm lag,i}$, trying to find any possible clues for the classification.

Spectral time lag is the time arrival difference between different energy bands for the prompt emission light curves of GRBs. Spectral lag was first introduced by \citet{Norris1996}. A cross-correlation function (CCF) can be used to quantify such an effect since the pulse peaks at different energy bands are delayed. The method is widely used to calculate spectral lag \citep{Ukwatta2010}. {A} positive spectral lag is {when} the high-energy photons {arrive} before the low-energy {ones}, a negative spectral lag is on the{opposite}. Prompt intrinsic spectral evolution or the curvature effect of relativistically moving shocked shells is used to explain the observed spectral lag \citep{Dermer2004, Uhm2016}.


Spectral time lag has been found correlated to several other quantities. The bolometric peak luminosity and spectral lag have an anti-correlation found by \citet{Norris2000}, later confirmed by \citet{Norris2002, Gehrels2006, Ukwatta2010}. Therefore, the lag is an indicator of both GRB peak luminosity and time history morphology, with short-lag and variable bursts having greater luminosities than long-lag and smooth bursts \citep{Norris2002, Ukwatta2012, Shao2017}. The peak luminosity and spectral lag relation can be viewed as being closely related to the peak luminosity and the variability relation. Variability should be inversely proportional to spectral lag, i.e., larger variability jets exhibit shorter spectral lags \citep{Zhang2009}. Using the internal shock model, the peak luminosity - spectral lag relation and the peak luminosity - variability relation can be caused by changes in observer's viewing angle with respect to the jet axis \citep{Ioka2001}. Bright GRBs are expected to have larger Lorentz factor and smaller viewing angles, meaning that the observer is viewing the GRB jet on axis. The spectral lag should be small due to the smaller emitting region \citep{Ioka2001}. \citet{Chen2005} showed the distribution of spectral time lags in GRBs. The distribution of spectral time lags in LGRBs is apparently different from SGRBs, which implies different physical mechanism. \citet{Yi2006} confirmed this result. \citet{Zhang2006} also studied the spectral time lag of SGRBs, and found the lags of the majority of SGRBs are so small that they are negligible or not measurable. \citet{Shao2017} carried out a {systematic} study of the spectral time lag properties of 50 single-pulsed GRBs detected by Fermi. \citet{Shao2017} provided a new measurement which is independent on energy channel selections, and the new results would favor the relativistic geometric effects for the origin of spectral time lag. \citet{Lu2018} found the spectral time lags are closely related to spectral evolution within the pulse. However, all of the statistics related to the lags do not have very high significance. In this work, we try to find if there are outliers, and to see whether the statistics become more significant without the outliers.
{We want to find more reliable correlations than the previous works. We hope to use the reliable correlations as standard candle and to constrain the cosmological parameters.}

This paper is organized as follows. Section 2 outlines the statistical methods. Section 3 discusses the PAM results. Section 4 is the conclusion and discussions.

\section{Statistical methods} \label{sec:method}

The parameters and samples are from our previous work \citep{Wang2020}. We collected all the possible data for 6289 GRBs in a big catalog \citep{Wang2020}, { of which 165 GRBs have been selected as they contain the required parameters}. The parameters we used in this work include $\tau_{\rm lag,i}$ (rest-frame spectral time lag, in unit of $\rm ms ~ MeV^{\rm -1}$), $T_{\rm 50,i}$ (duration of 25\% to 75\% $\gamma$-ray fluence in rest-frame), $T_{\rm 90,i}$ (duration of 5\% to 95\% $\gamma$-ray fluence in rest-frame), $T_{\rm R45,i}$ \citep[defined in][]{Reichart2001}, $variability_{\rm 2}$ (light curve variability from the definition of \citet{Reichart2001}), $L_{\rm pk,52}$ (peak luminosity of 1 $\rm s$ time bin in rest-frame 1-$10^{4}$ $\rm keV$ energy band,  in unit of $\rm 10^{\rm 52} ~ erg ~ s^{\rm -1}$), $E_{\rm iso,52}$ (isotropic $\gamma$-ray energy in rest-frame 1-$10^{4}$ $\rm keV$ energy band,  in unit of $\rm 10^{\rm 52} ~ ergs$), {$\alpha_{\rm Band}$} (low energy spectral index of {Band} model), {$\beta_{\rm Band}$} (high energy spectral index of {Band} model), {$E_{\rm p,Band,i}$} (rest-frame spectral peak energy of {Band} model, $\rm keV$), $E_{\rm p,i}$ (rest-frame spectral peak energy of {Band} model and cutoff power law model, $\rm keV$), $\alpha_{\rm cpl}$ (low energy spectral index of cutoff power law model), $\log t_{\rm burst,i}$ (rest-frame central engine active duration in logarithm,  in unit of $\rm s$) \citep[defined in][]{Zhang2014}, $t_{\rm radio,pk,i}$ (rest-frame peak time in radio band,  in unit of $\rm s$), $\beta_{\rm X11hr}$ (index in X-ray band at 11 hours related to the trigger time), Age ( in unit of $\rm Myr$), $A_{\rm V}$ (dust extinction), host galaxy offset (the distance from GRB location to the centre of its host galaxy, in unit of $\rm kpc$), Mag (absolute magnitude in AB system at rest 3.6 $\mu m$ wavelength), $\log SSFR$ (specific star formation rate in logarithm,  in unit of $\rm Gyr^{\rm -1}$), $N_{\rm H}$ (column density of hydrogen,  in unit of $\rm 10^{\rm 21} ~ cm^{\rm -2}$). We use label ``i" to mark the parameters in rest-frame. We choose three parameters including $\tau_{\rm lag,i}$ for PAM analysis. {For the spectral time lag, because the energy band is different for different instrument, we divide the spectral time lag by the difference of two energy band central values to get a unified quantity, which can be seen in table 1 of \citet{Wang2020}. For example, the spectral time lag for GRB 980425B is 1.46 $\pm$ 0.18 $\rm s$ between 50-100 $\rm keV$ and 25-50 $\rm keV$ \citep{Zhang2009}, and the redshift is 0.0085. The central value of $\tau_{\rm lag,i}$ is $\rm \frac{1.46}{1+0.0085} \cdot (\frac{50+100}{2} - \frac{25+50}{2}) \cdot 10^{6} = 38605 ms ~ MeV^{\rm -1}$. The error of $\tau_{\rm lag,i}$ is $\rm \frac{0.18}{1+0.0085} \cdot (\frac{50+100}{2} - \frac{25+50}{2}) \cdot 10^{6} = 4760 ms ~ MeV^{\rm -1}$.} Same as the suggestion in \citet{Foley2008}, we removed the $\tau_{\rm lag,i}$ for GRB 060218,  {because GRB 060218 has extremely large lag, and it is an X-ray flash rather than a typical GRB.}

For the spectral parameters, the spectra are mainly fitted by three models: Band model, cutoff power law (CPL) model and simple power law (SPL) model \citep{Li2016}. Band model is a smoothly joint broken power law with the definition \citep{Band1993}:
\begin{equation}
\label{eq:Band}
N(E)=
\left.
\Big \{
\begin{array}{lr}
A\left(\frac{E}{100\ {\rm keV}}\right)^{\alpha} e ^{-\frac{E}{E_0}},& E < (\alpha-\beta)E_0, \\
A\left(\frac{E}{100\ {\rm keV}}\right)^{\beta} \left[ \frac{(\alpha-\beta)E_0}{100\ {\rm keV}}\right]^{\alpha-\beta} e^{\beta-\alpha},& E \ge (\alpha-\beta)E_0,
\end{array}
\right.
\end{equation}
where $\alpha$ is low energy photon index, $\beta$ is high energy photon index, $A$ is the coefficient for normalization, $E$ is the energy of the photons, and $E_0$ is the break energy. Mostly we used $E_{\rm p}$ instead of $E_0$. $E_{\rm p}$ is the peak energy in spectrum of $E^2N$, and $E_{\rm p}=(2+\alpha)E_0$.

In the previous work \citep{Wang2020}, we use {$\alpha_{\rm Band}$}, {$\beta_{\rm Band}$} and {$E_{\rm p,Band}$} as Band function spectral parameters. $\alpha_{\rm cpl}$ and $E_{\rm p,cpl}$ to mark CPL model spectrum parameters. The formula of SPL model is $N(E) = A E^{\alpha_{\rm spl}}$. {In this paper}, we use {$- \alpha_{\rm Band}$}, {$- \beta_{\rm Band}$} and $- \alpha_{\rm cpl}$ to stand for the opposite of spectral indices, { as they become mostly positive numbers by adding the `$-$' sign}.

Cluster analysis divides data into clusters that are meaningful, useful, or both. Cluster analysis is the study of techniques for finding the most representative cluster prototypes. A cluster is a set of objects in which each object is more similar to the prototype that defines the cluster than to the prototype of any other clusters. There are a number of such techniques, but two of the most prominent are K-means and K-medoids. K-medoids method is more robust than others in terms of outliers. {The outliers are the smallest groups with a few points in this paper.} Partitioning Around Medoids (PAM) is the most prominent method of K-medoids. In this paper, we use PAM method for outlier detection. 
In order to avoid the influence of different unit, we apply data standardization to all the parameters. We use the R language function Nbclust to calculate the best cluster number.
The detailed processes are the following:
\begin{itemize}
\item We calculate the best cluster number K and choose K initial centroids, where K is a user-specified parameter, namely, the number of clusters desired. In this paper ,the K is 2 or 3;
\item We use the {Euclidean} distance as {similarity measurement}. We calculate the {Euclidean} distance to the initial centroids of each point;
\item Each point is then assigned to the closest centroid, and each collection of points assigned to a centroid is a cluster;
\item For every cluster, we calculate the euclidean distance sum to the centroid of each point, which are assigned to this centroid. Then we get the sum of each cluster;
\item For every cluster, we choose one point to update the centroid;
\item Points are assigned to the updated centroids;
\item For every cluster, we calculate the euclidean distance sum to the updated centroid of each point, which are assigned to this updated centroid. Then we get the sum of all the clusters;
\item If the sum of all the clusters in the seventh step is smaller than the fourth step, we update the centroids;
\item We repeat the assignments and update steps until no point changes clusters, or equivalently, until the centroids remain the same.
\end{itemize}

{We tried all the three parameters combinations including $\tau_{\rm lag,i}$. For every combination, we use PAM method for outlier detection firstly. If we find outliers, we do the linear regression among the three parameters including $\tau_{\rm lag,i}$ without outliers. Some combinations have obvious outliers and significant correlations as shown in Section \ref{subsec:regression}. Some combinations just have obvious outliers, but no significant {correlations can be found, as} shown in Section \ref{subsec:outliers}. Some combinations have no obvious outliers and no significant correlations, and we don't show these results.} We also consider all the error bars using MC method \citep{Zou2017}. The results are shown in the next section.

\section{PAM results} \label{sec:result}

We tried all the three combinations including $\tau_{\rm lag,i}$ with PAM method. The data selection criterion is that the sample should have at least 10 GRBs. We use the R language Nbclust package to calculate the best cluster number. Nbclust package provides 30 indices for determining the number of clusters, like Krzanowski-Lai (KL) index \citep{Dudoit2002}, Davies-Bouldin (DB) index \citep{Davies1979}, and so on. {\citet{Charrad2014} showed more details about Nbclust package.} As shown in the upper panel of Figure \ref{fig:beta_bandT50ispectral_lagi}, number 2 has the maximum criteria. Therefore, the cluster number should be 2. However, the cluster number of Figures \ref{fig:beta_bandMagspectral_lagi} and \ref{fig:beta_bandspectral_lagilog_t_bursti} is 3, as shown in Figure \ref{fig:beta_bandspectral_lagilog_t_bursti}. Except Figures \ref{fig:beta_bandMagspectral_lagi} and \ref{fig:beta_bandspectral_lagilog_t_bursti}, the cluster number of other figures is 2 with Nbclust package. We did not show the others in this paper. We found 191 combinations with obvious outliers. Though PAM method can find the outliers, the rest of the sample in the PAM plot does not {show} any meaningful message on the correlation or the clustering. Therefore, some independent statistics should be {applied} to check the statistical property for the {rest of the sample} Linear regression is carried out with outliers removed. Only 8 combinations have passed the hypothesis testing, which are displayed in Figures \ref{fig:beta_bandT50ispectral_lagi} to \ref{fig:variability2Epispectral_lagi}. However for the passed combinations, the correlation is not very significant. {We use the adjusted $R^{2}$ to measure the goodness of the regression model. It means the percentage of variance explained considering the parameter freedom.} We also showed a small part of cluster plots with obvious outliers, but not passing the hypothesis testing in Figures \ref{fig:beta_bandspectral_lagilog_t_bursti} to \ref{fig:offsetlog_SSFRspectral_lagi}. These figures include all the outliers. In other words, the cluster plots with repeated outliers are not displayed in this paper.


\subsection{Remarkable linear regression results without outliers} \label{subsec:regression}

\begin{figure}
\centering
\includegraphics[width=0.45\textwidth]{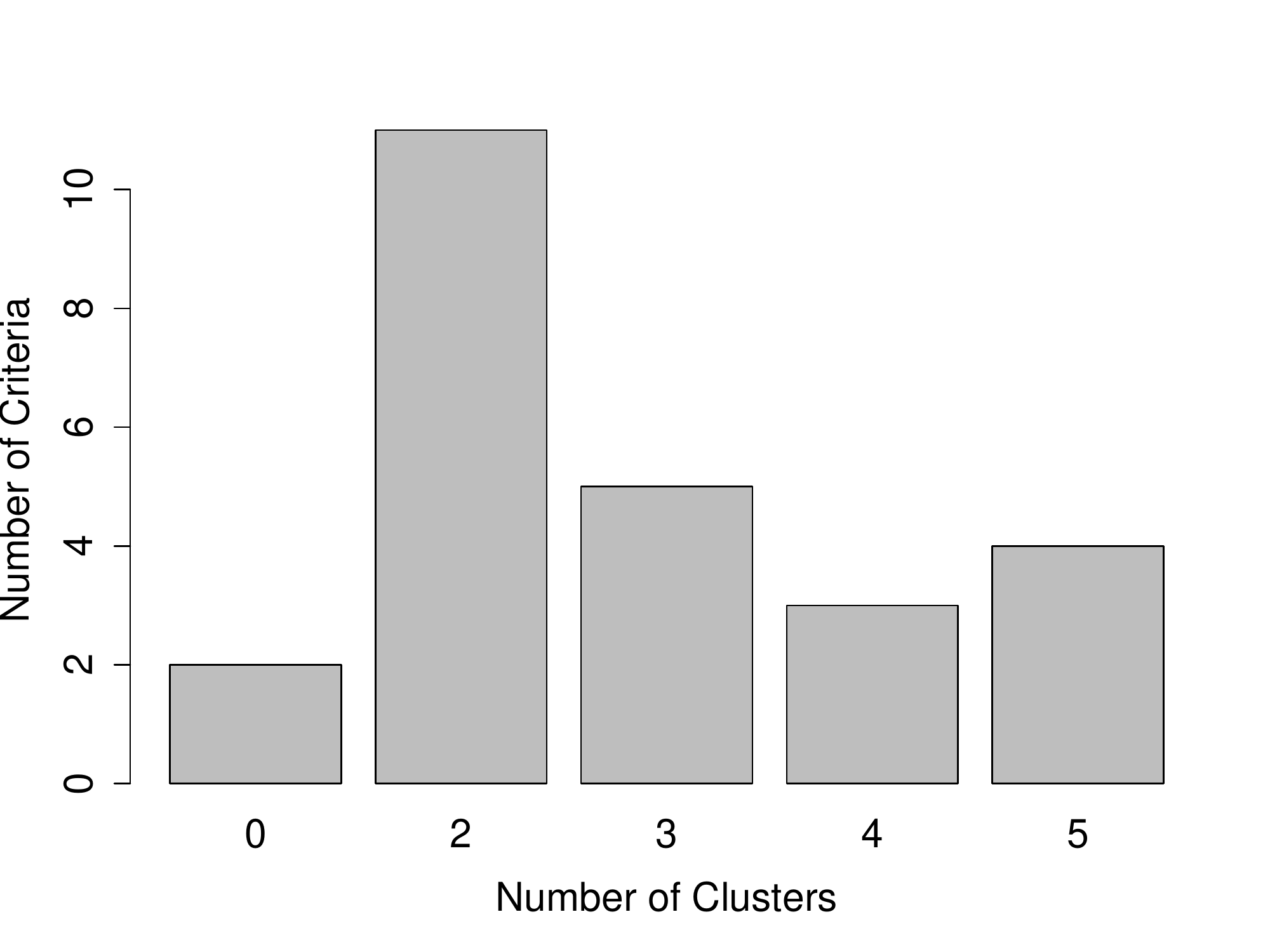}

\includegraphics[width=0.45\textwidth]{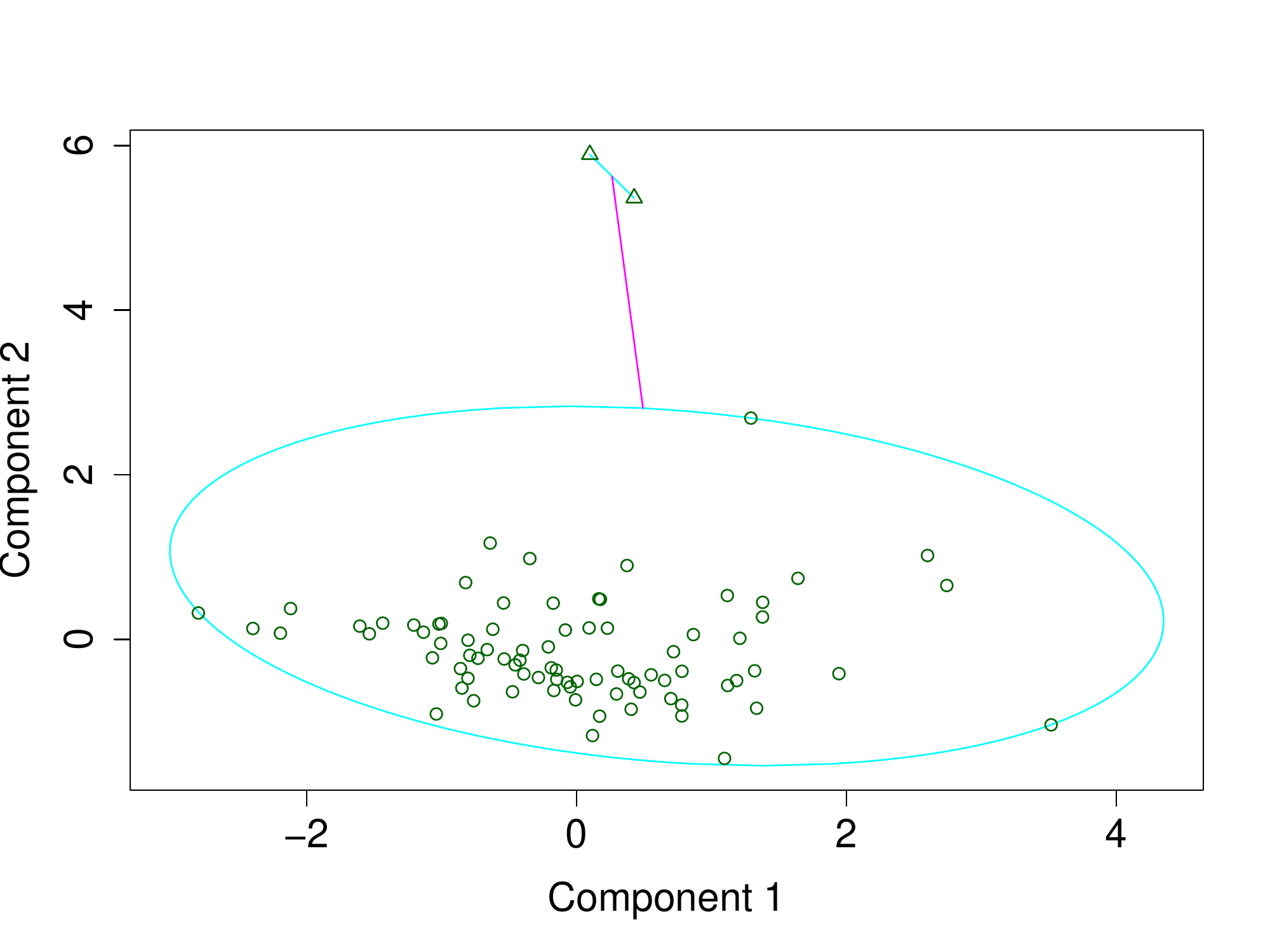}
\includegraphics[width=0.45\textwidth]{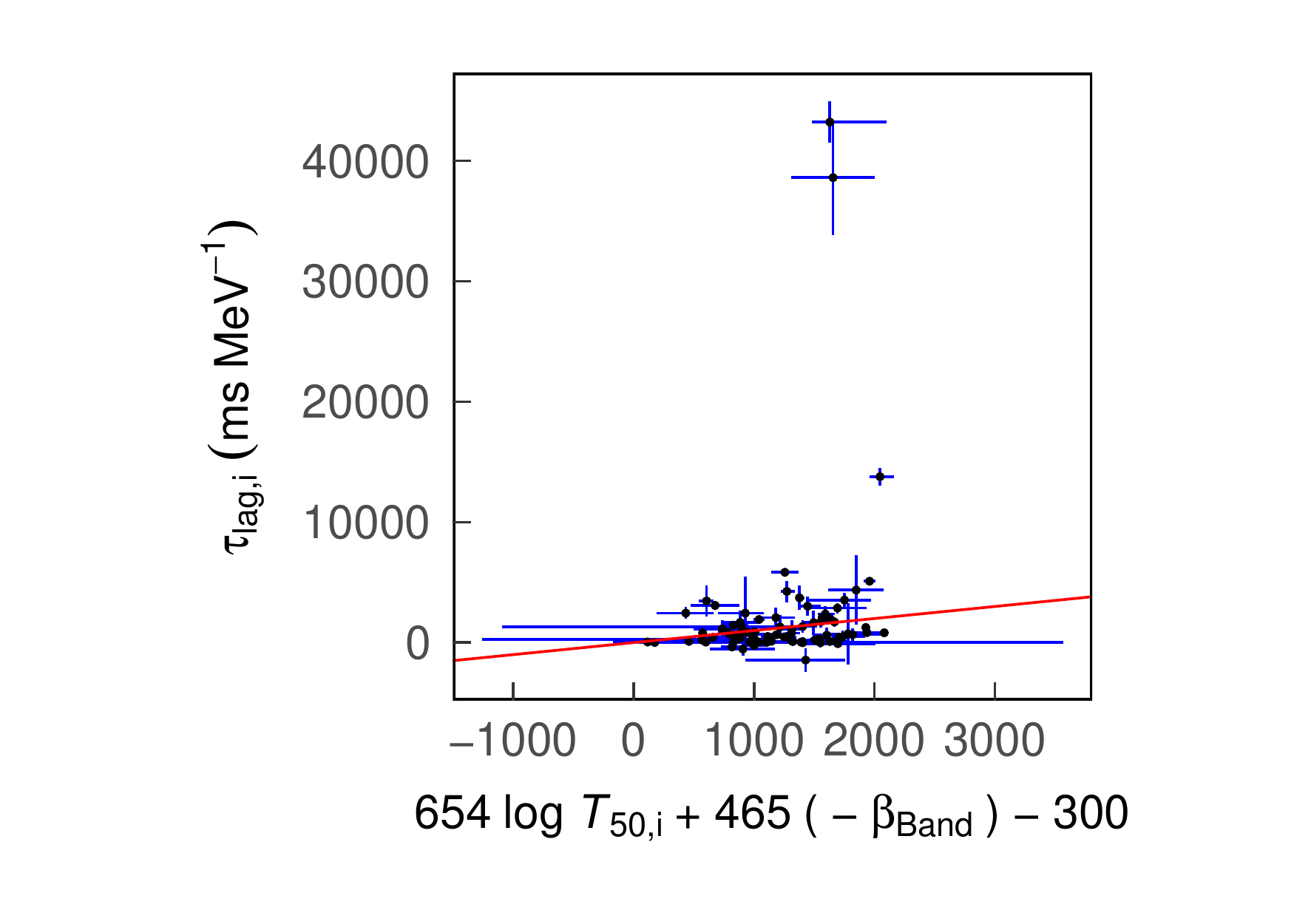}
\caption{The outlier plot for the three parameters, $\tau_{\rm lag,i}$, $T_{\rm 50,i}$ and {$\beta_{\rm Band}$}. The histogram is the result of function NbClust, indicating the best cluster number is 2. The left panel shows the PAM result for the clustering analysis. Notice the x and y axes (compnent 1 and 2) {are the two principal components in the PAM method and} do not represent any combination of the three parameters. The right plot shows the regression result of the physical parameters without outliers. The outliers are GRBs 980425B and 030528A. The linear regression result is $\tau_{\rm lag,i} = (654^{\rm +150}_{\rm -150}) \times \log T_{\rm 50,i} + (465^{\rm +400}_{\rm -270}) \times (-\beta_{\rm Band}) - (300^{\rm +400}_{\rm -940})$. The adjusted $R^{2}$ is 0.07. The red line is the best fit result, i.e., $y = x$.}
\label{fig:beta_bandT50ispectral_lagi}
\end{figure}

\begin{figure}
\includegraphics[width=0.45\textwidth]{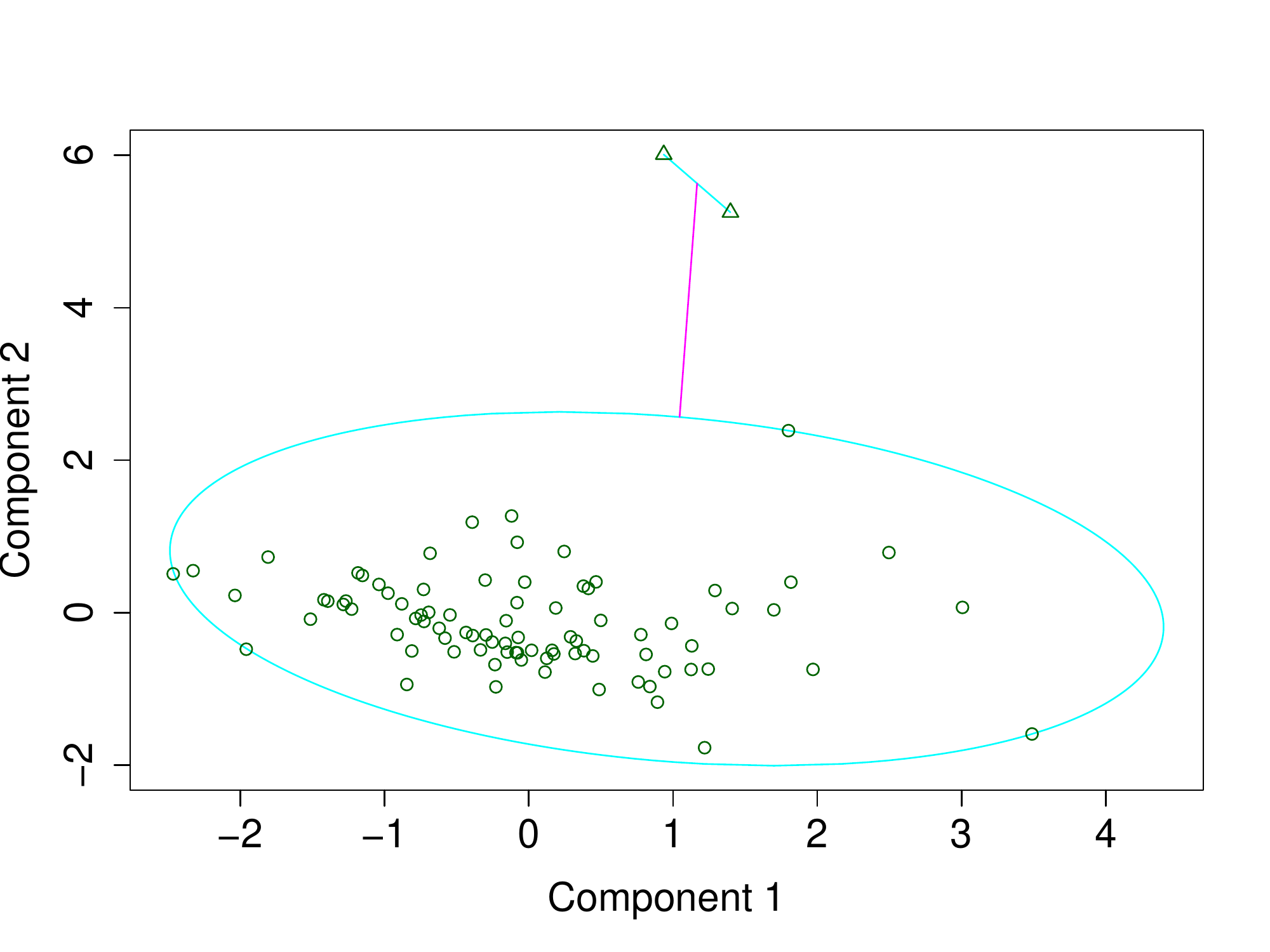}
\includegraphics[width=0.45\textwidth]{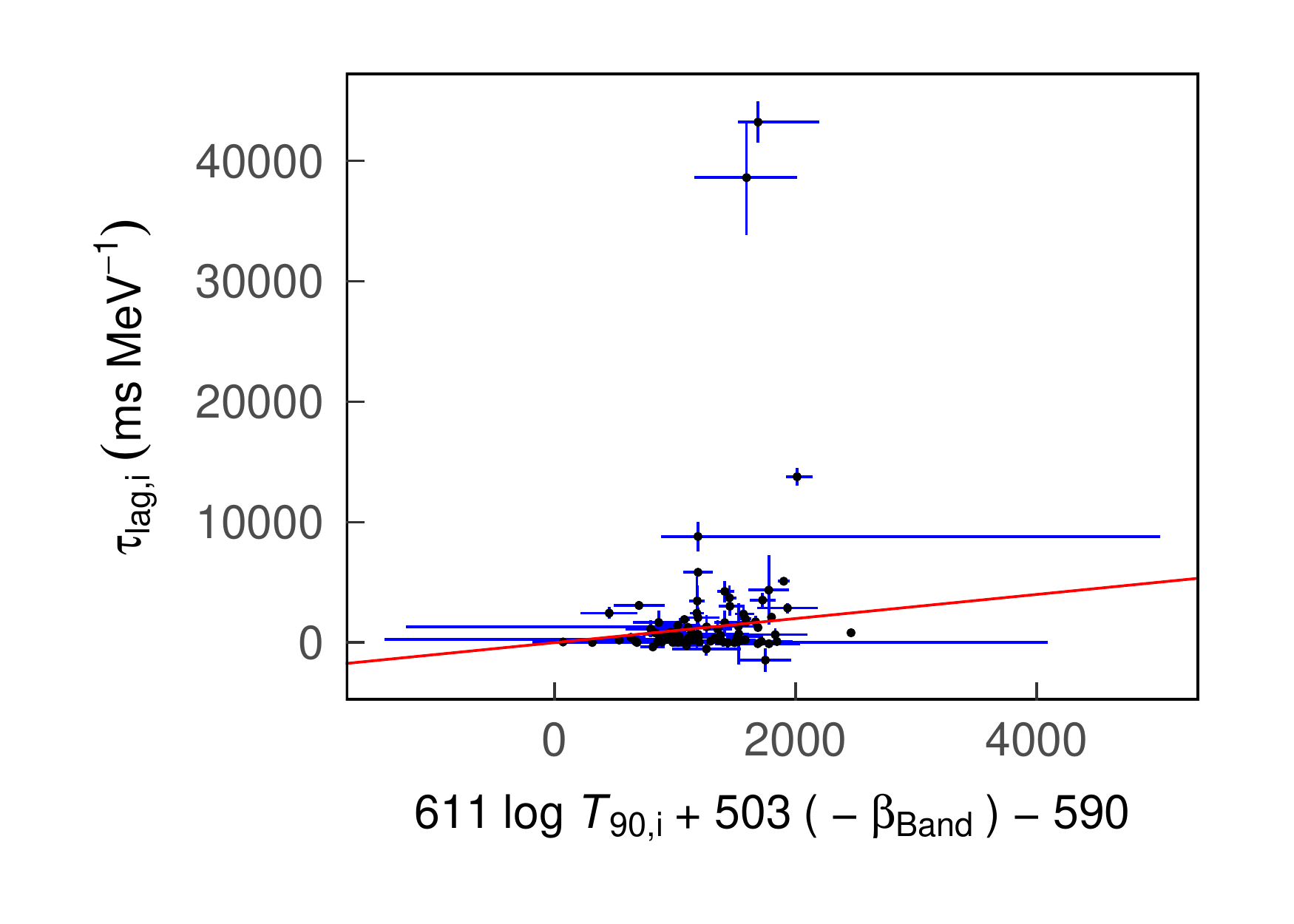}
\caption{Similar to Figure \ref{fig:beta_bandT50ispectral_lagi}, with the three parameters being $\tau_{\rm lag,i}$, $T_{\rm 90,i}$ and {$\beta_{\rm Band}$}. The outliers are GRBs 980425B and 030528A.  The linear regression result is $\tau_{\rm lag,i} = (611^{\rm +130}_{\rm -130}) \times \log T_{\rm 90,i} + (503^{\rm +430}_{\rm -310}) \times (-\beta_{\rm Band}) - (590^{\rm +440}_{\rm -1100})$. The adjusted $R^{2}$ is 0.05.} 
\label{fig:beta_bandT90ispectral_lagi}
\end{figure}

\begin{figure}
\includegraphics[width=0.45\textwidth]{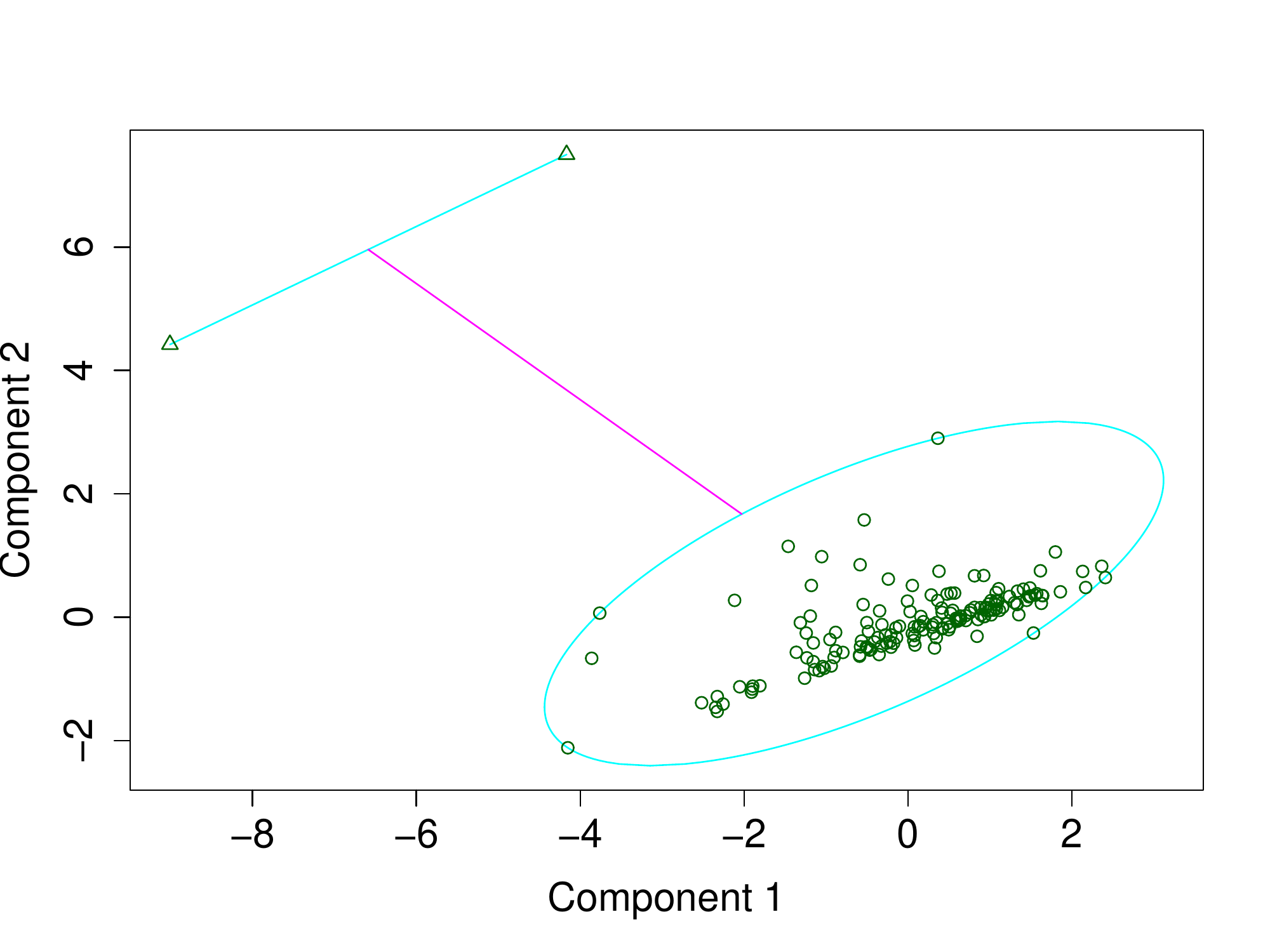}
\includegraphics[width=0.45\textwidth]{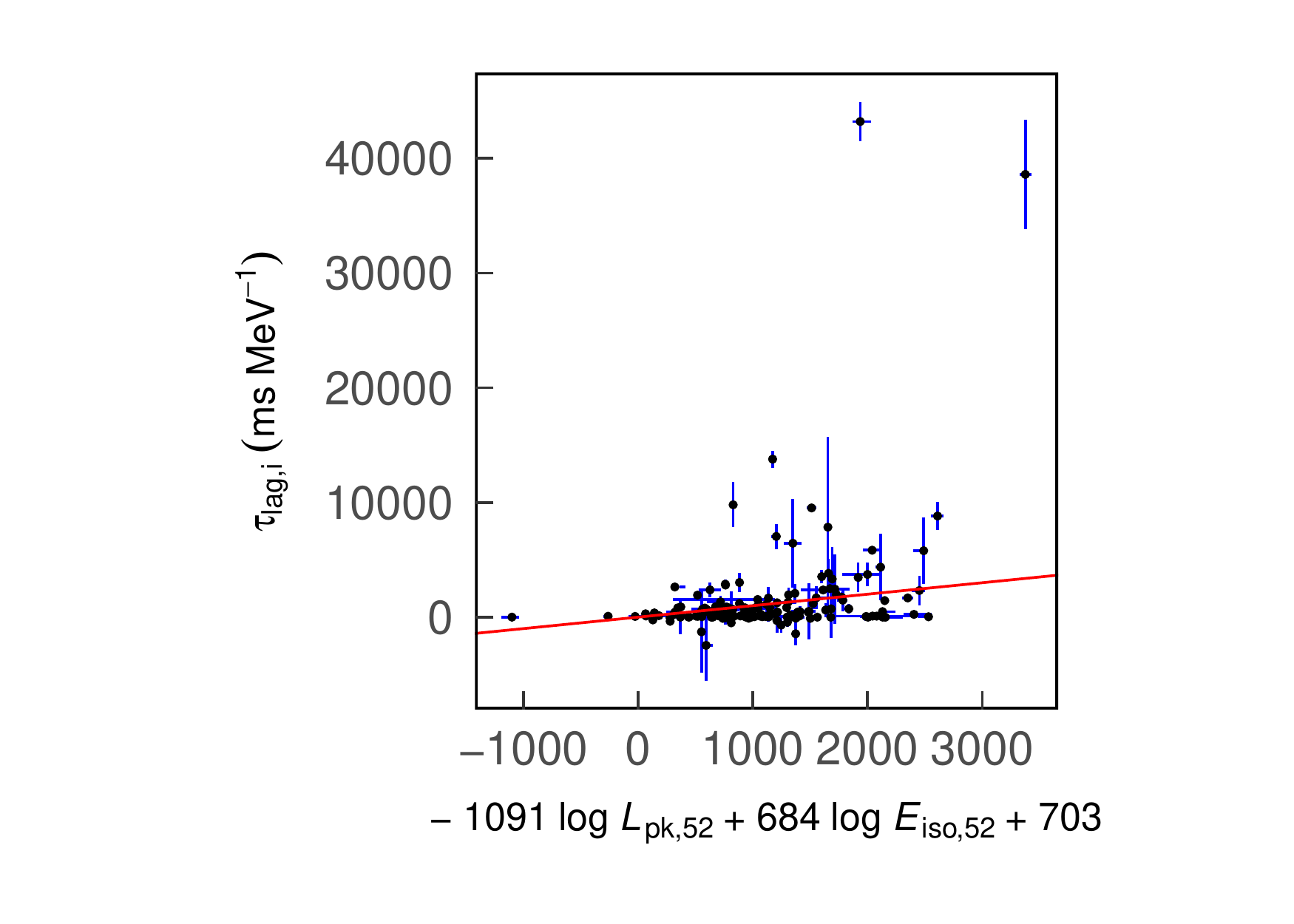}
\caption{Similar to Figure \ref{fig:beta_bandT50ispectral_lagi}, with the three parameters being $\tau_{\rm lag,i}$, $L_{\rm pk,52}$ and $E_{\rm iso,52}$. The outliers are GRBs 980425B and 030528A.  The linear regression result is $\tau_{\rm lag,i} = (-1091^{\rm +150}_{\rm -160}) \times \log L_{\rm pk,52} + (684^{\rm +120}_{\rm -130}) \times \log E_{\rm iso,52} + (703^{\rm +37}_{\rm -85})$. The adjusted $R^{2}$ is 0.08.}
\label{fig:E_isoL_pkspectral_lagi}
\end{figure}

\begin{figure}
\includegraphics[width=0.45\textwidth]{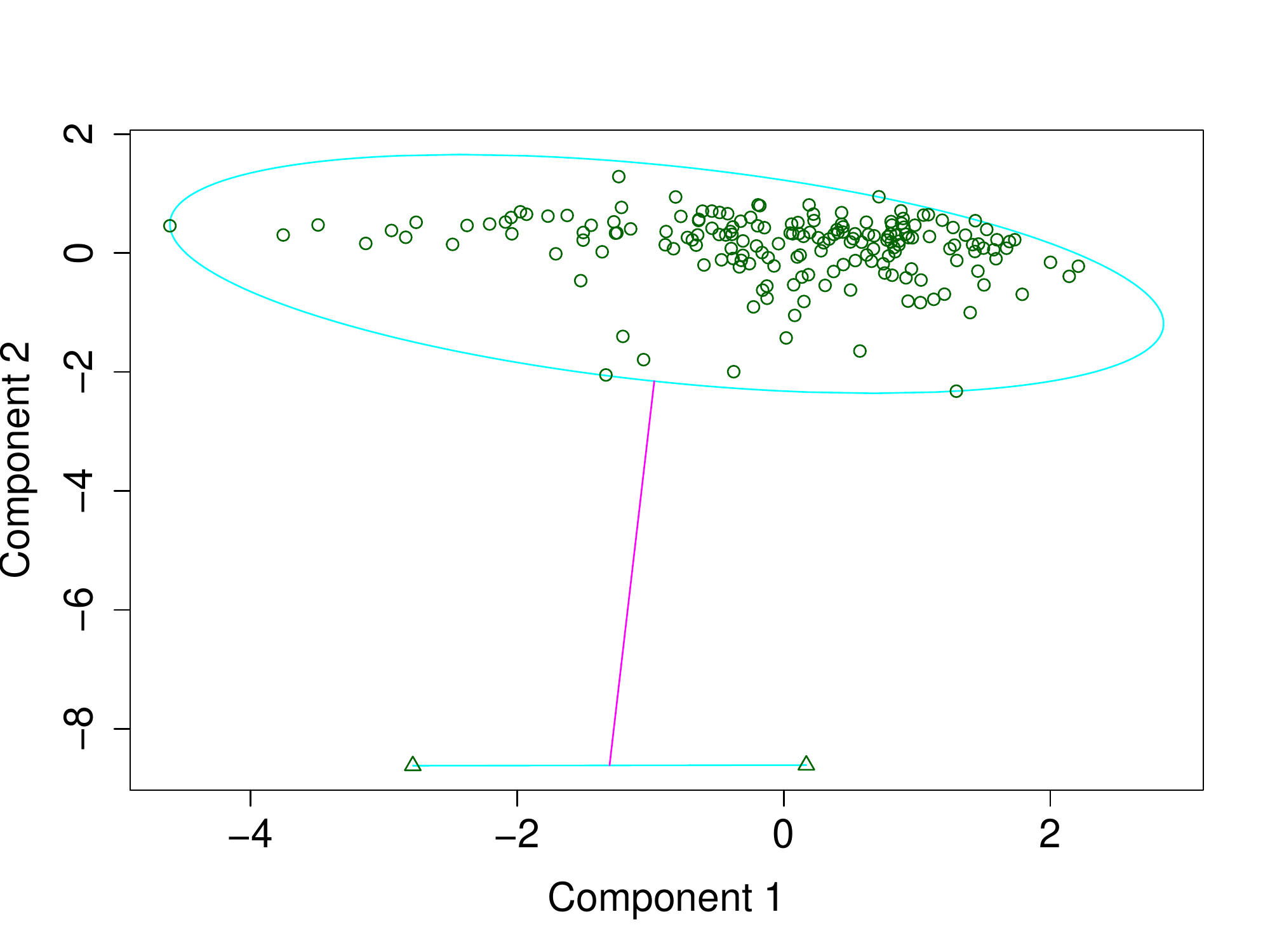}
\includegraphics[width=0.45\textwidth]{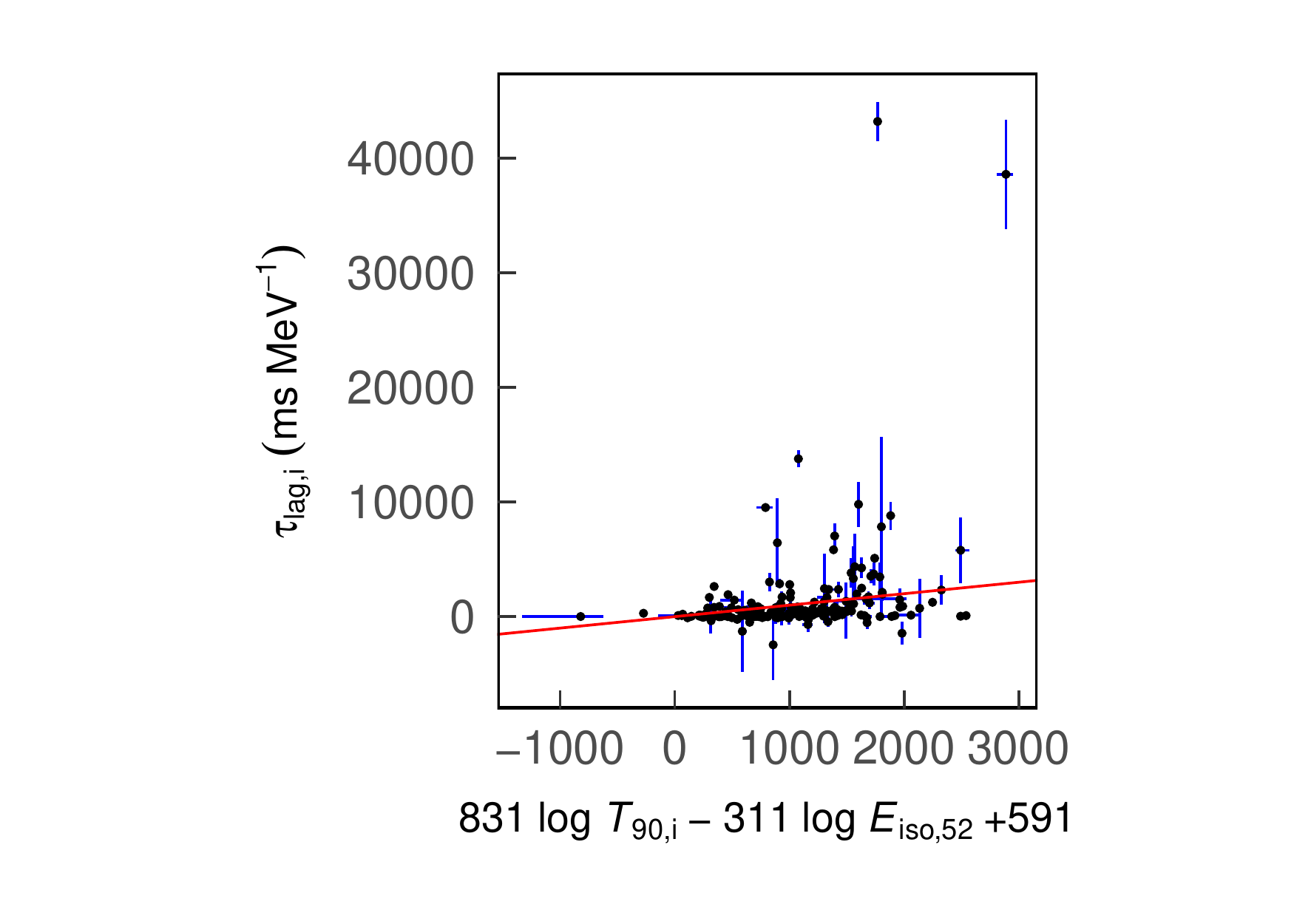}
\caption{Similar to Figure \ref{fig:beta_bandT50ispectral_lagi}, with the three parameters being $\tau_{\rm lag,i}$, $T_{\rm 90,i}$ and $E_{\rm iso,52}$. The outliers are GRB 980425B and 030528A. The linear regression result is $\tau_{\rm lag,i} = (831^{\rm +126}_{\rm -115}) \times \log T_{\rm 90,i} - (311^{\rm +80}_{\rm -78}) \times \log E_{\rm iso,52} + (591^{\rm +24}_{\rm -58})$. The adjusted $R^{2}$ is 0.07.}
\label{fig:E_isoT90ispectral_lagi}
\end{figure}

\begin{figure}
\includegraphics[width=0.45\textwidth]{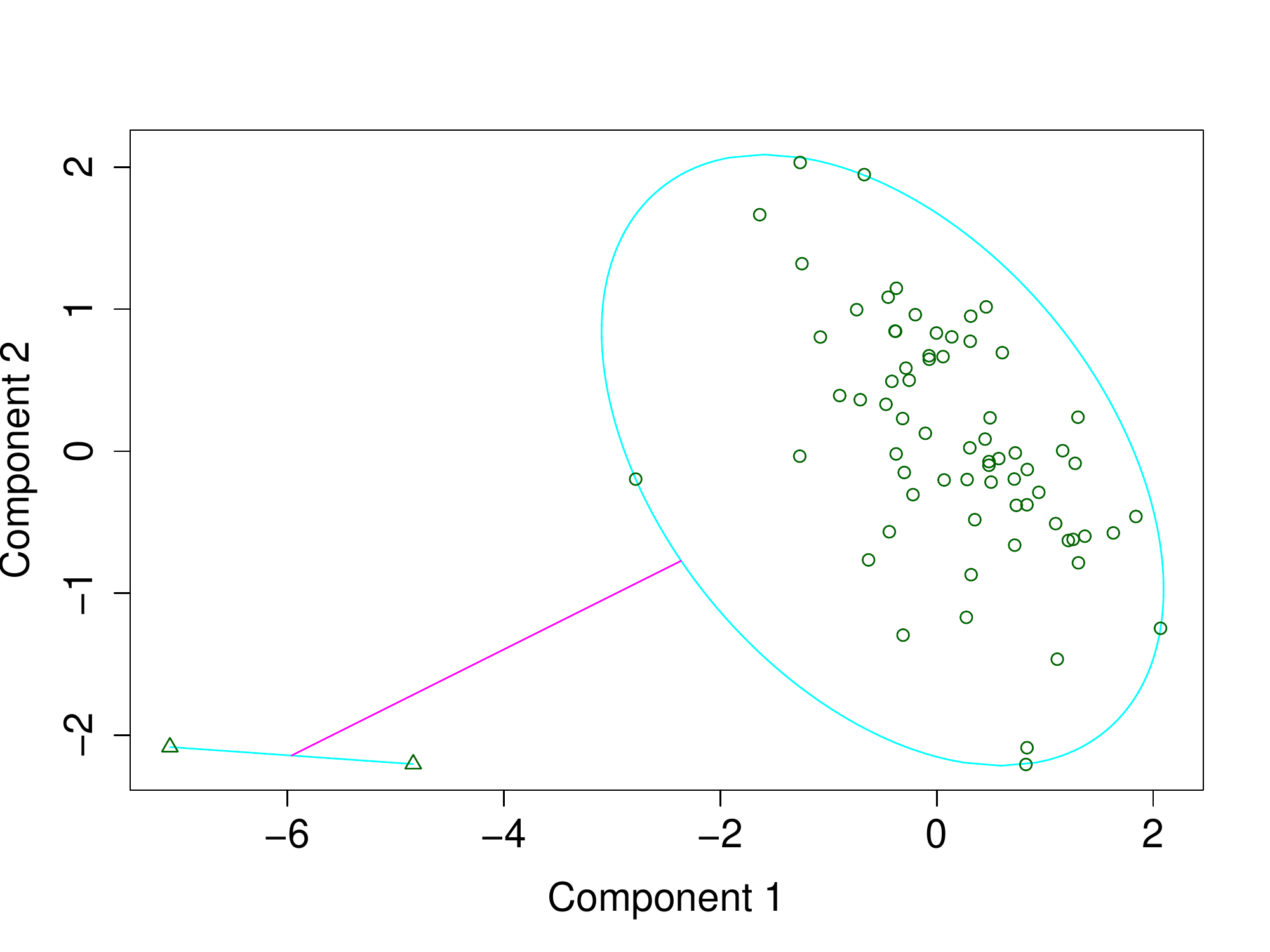}
\includegraphics[width=0.45\textwidth]{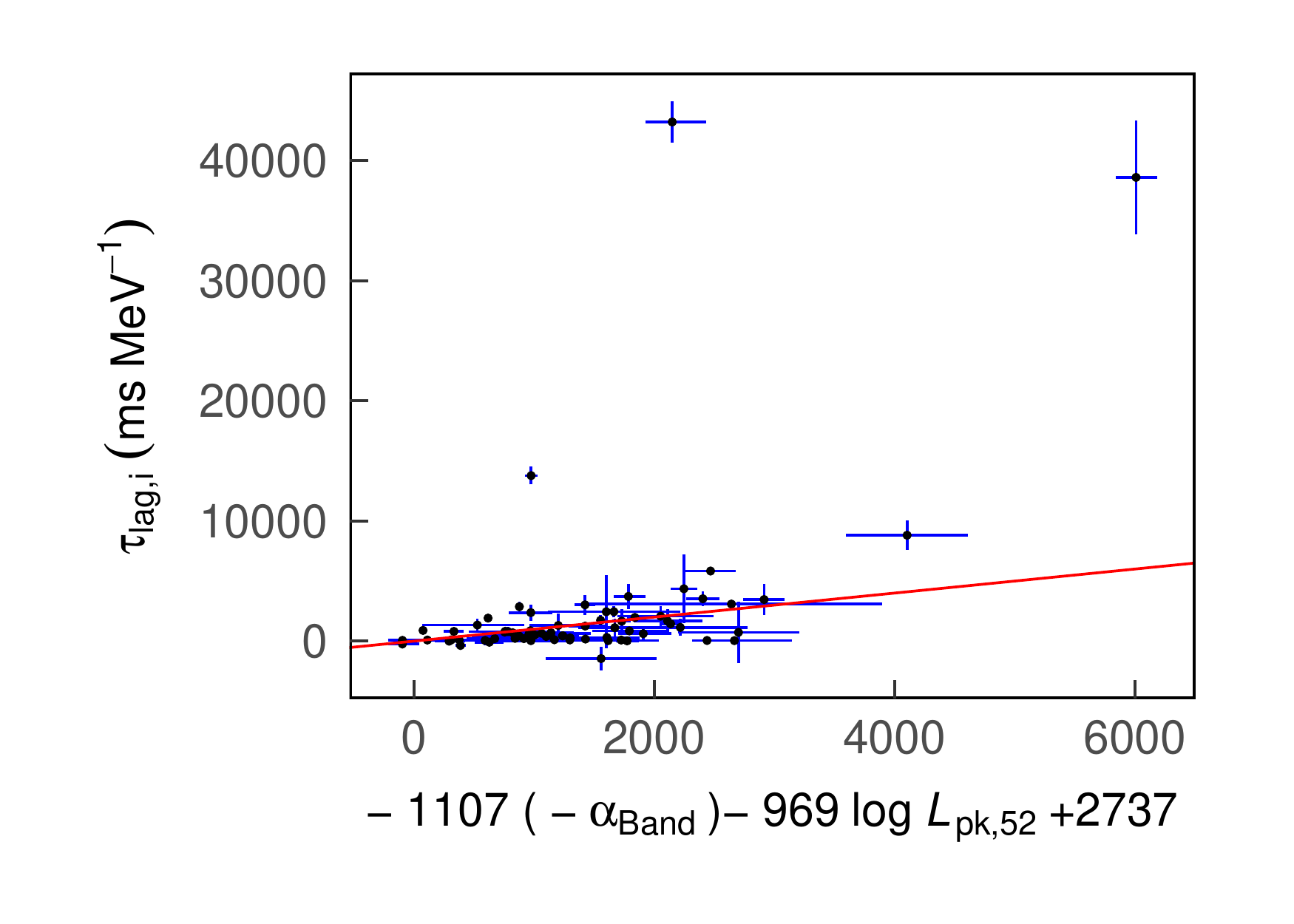}
\caption{Similar to Figure \ref{fig:beta_bandT50ispectral_lagi}, with the three parameters being $\tau_{\rm lag,i}$,  {$\alpha_{\rm Band}$} and $L_{\rm pk,52}$. The outliers are GRB 980425B and 030528A.  The linear regression result is $\tau_{\rm lag,i} = (-1107^{\rm +520}_{\rm -670}) \times (-\alpha_{\rm Band}) - (969^{\rm +150}_{\rm -150}) \times \log L_{\rm pk,52} + (2737^{\rm +230}_{\rm -570})$. The adjusted $R^{2}$ is 0.14.}
\label{fig:L_pkalpha_bandspectral_lagi}
\end{figure}

\begin{figure}
\includegraphics[width=0.45\textwidth]{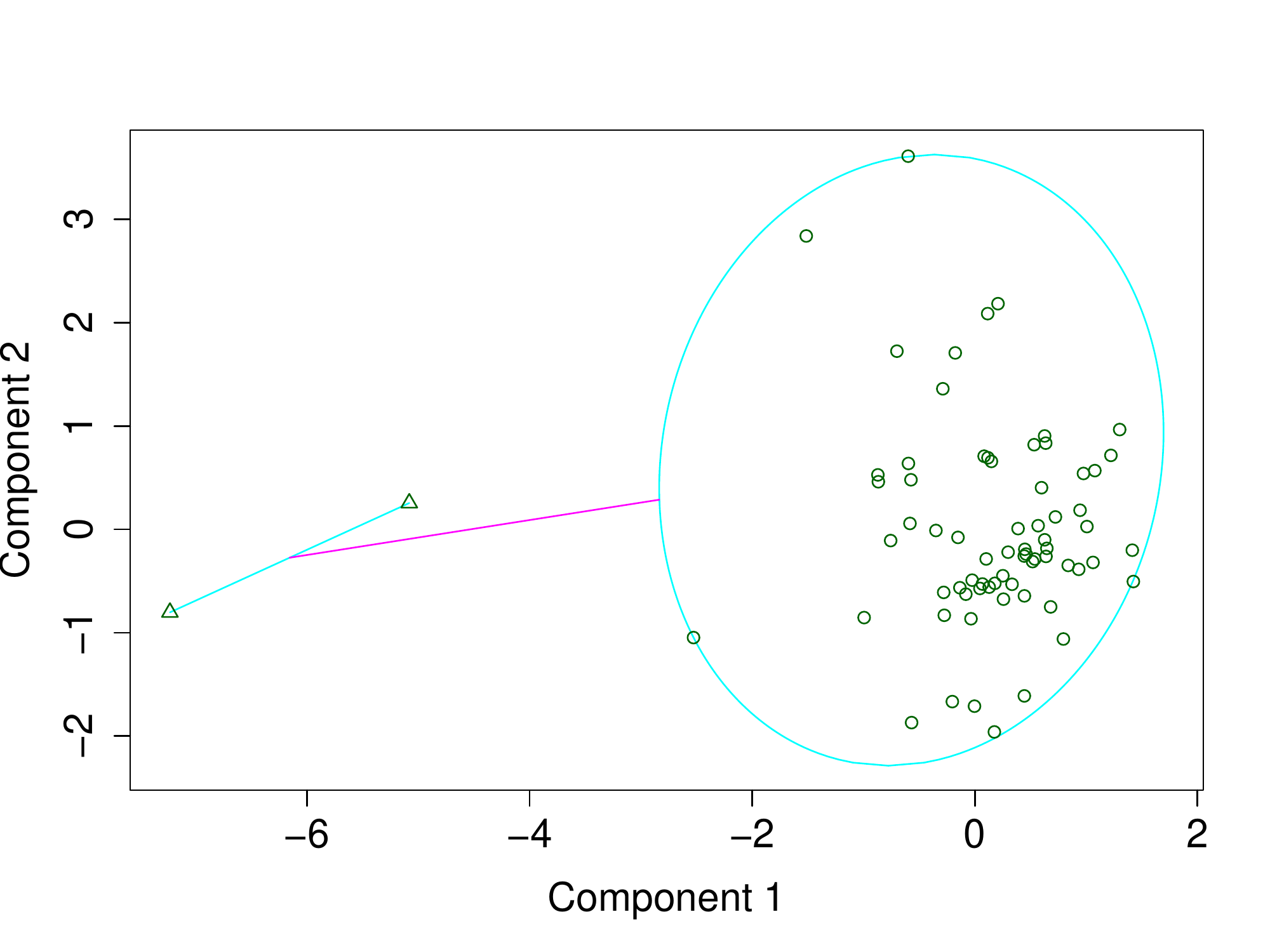}
\includegraphics[width=0.45\textwidth]{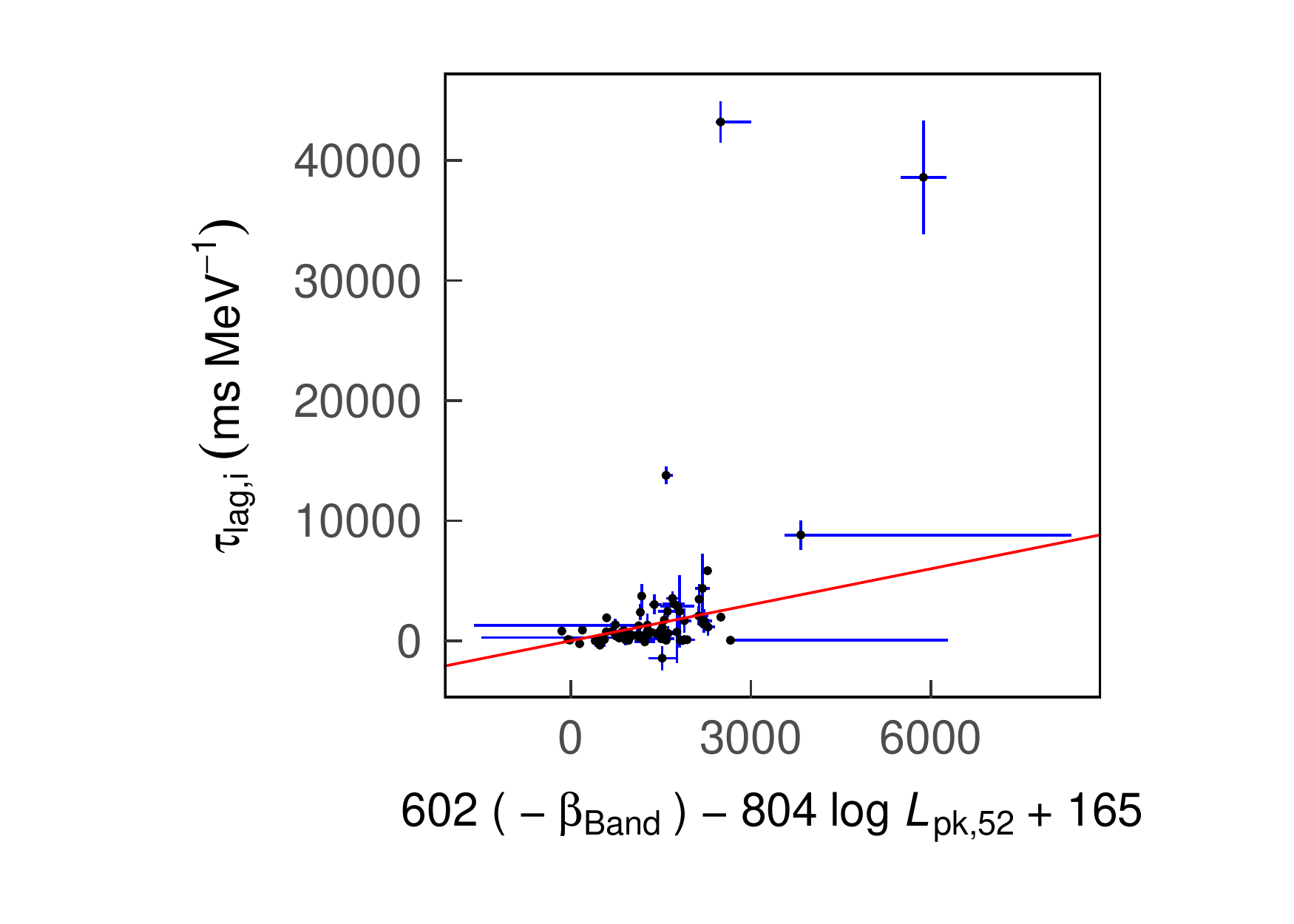}
\caption{Similar to Figure \ref{fig:beta_bandT50ispectral_lagi}, with the three parameters being $\tau_{\rm lag,i}$,  {$\beta_{\rm Band}$} and $L_{\rm pk,52}$. The outliers are GRB 980425B and 030528A.  The linear regression result is $\tau_{\rm lag,i} = (602^{\rm +640}_{\rm -370}) \times (-\beta_{\rm Band}) - (804^{\rm +240}_{\rm -180}) \times \log L_{\rm pk,52} + (165^{\rm +520}_{\rm -1400})$. The adjusted $R^{2}$ is 0.21.}
\label{fig:L_pkbeta_bandspectral_lagi}
\end{figure}

\begin{figure}
\includegraphics[width=0.45\textwidth]{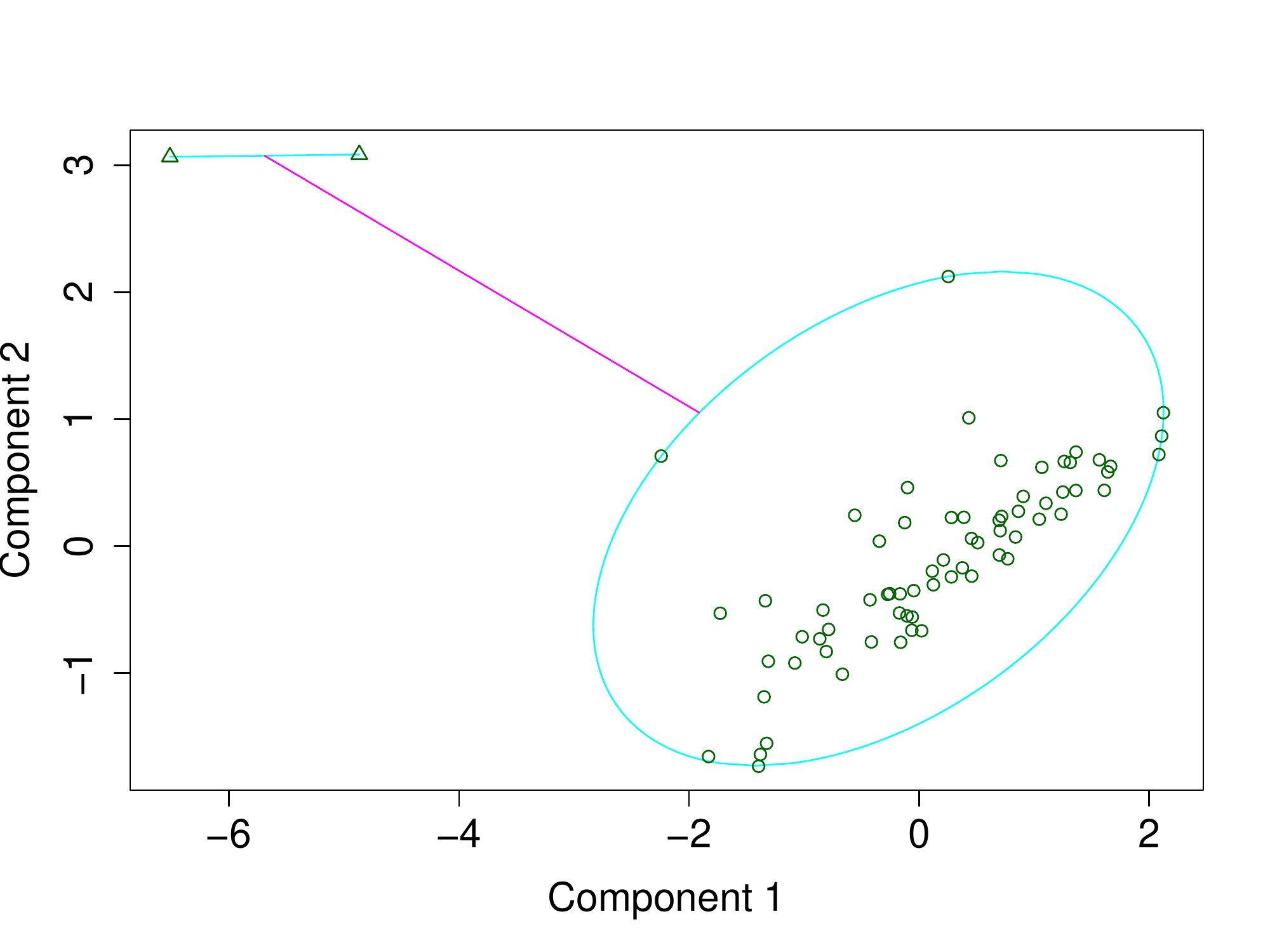}
\includegraphics[width=0.45\textwidth]{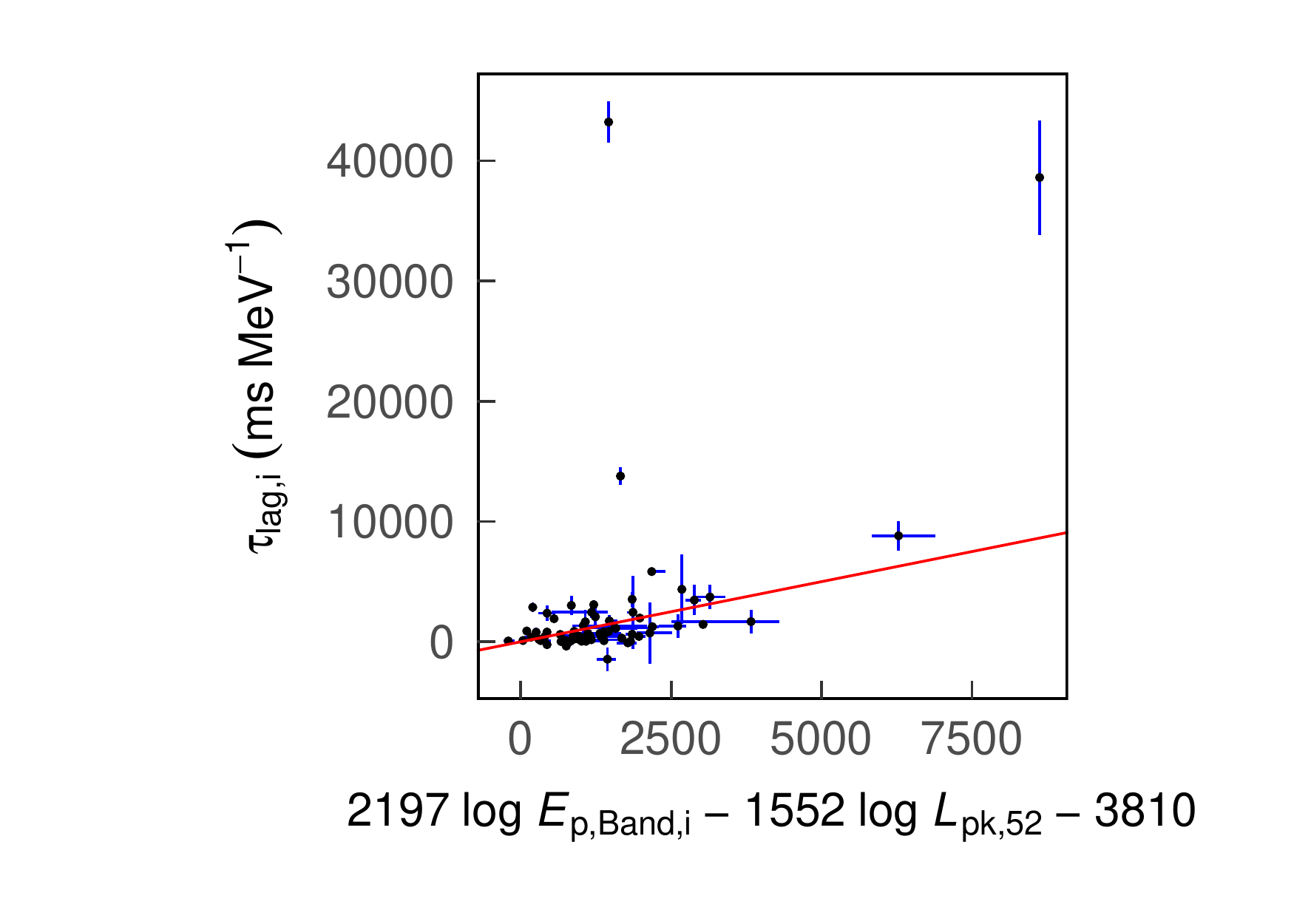}
\caption{Similar to Figure \ref{fig:beta_bandT50ispectral_lagi}, with the three parameters being $\tau_{\rm lag,i}$,  {$E_{\rm p,Band,i}$} and $L_{\rm pk,52}$. The outliers are GRB 980425B and 030528A.  The linear regression result is $\tau_{\rm lag,i} = (2197^{\rm +410}_{\rm -390}) \times \log E_{\rm p,Band,i} - (1552^{\rm +220}_{\rm -230}) \times \log L_{\rm pk,52} - (3810^{\rm +360}_{\rm -950})$. The adjusted $R^{2}$ is 0.22.}
\label{fig:L_pkE_P_bandispectral_lagi}
\end{figure}

\begin{figure}
\includegraphics[width=0.45\textwidth]{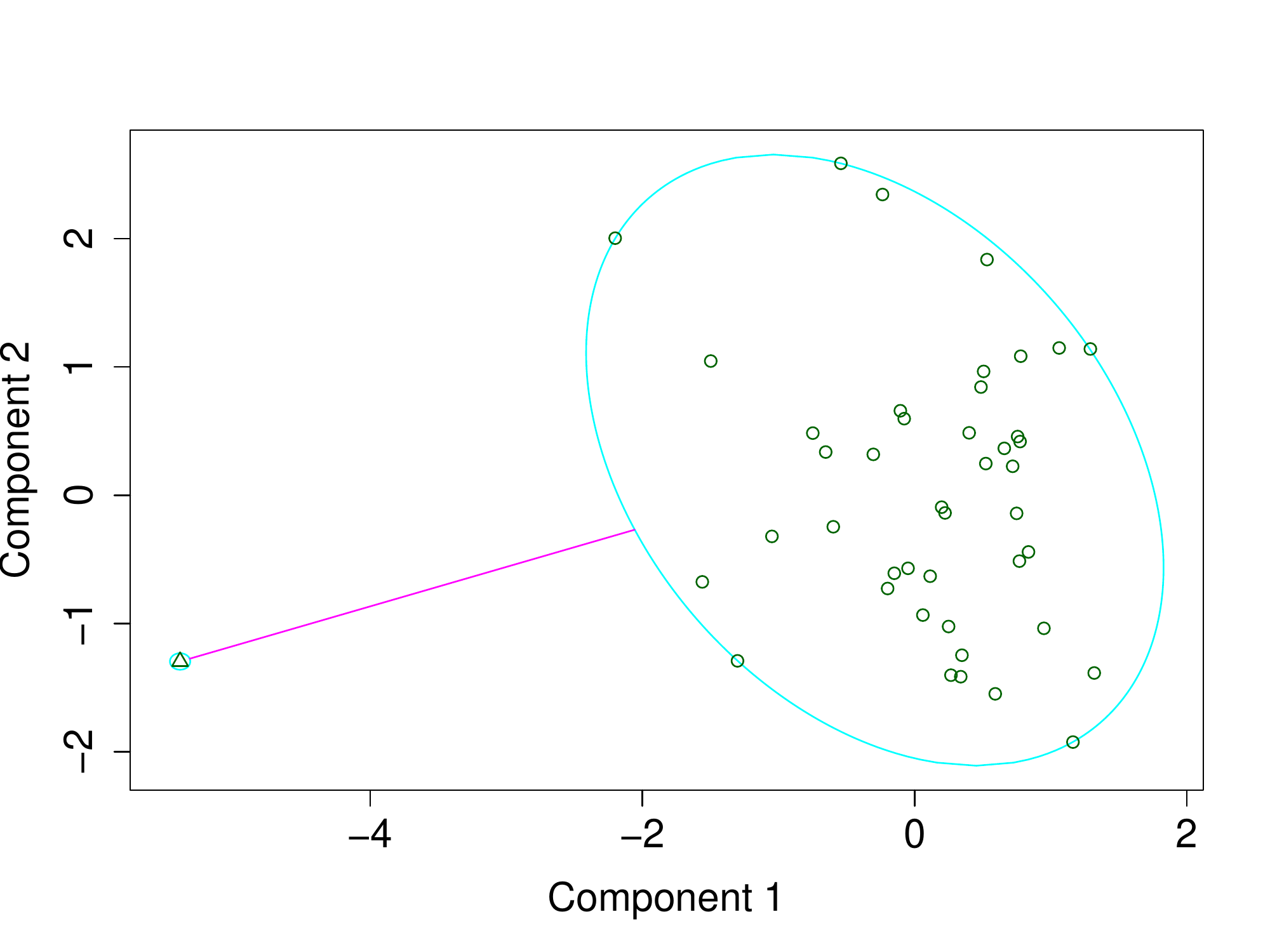}
\includegraphics[width=0.45\textwidth]{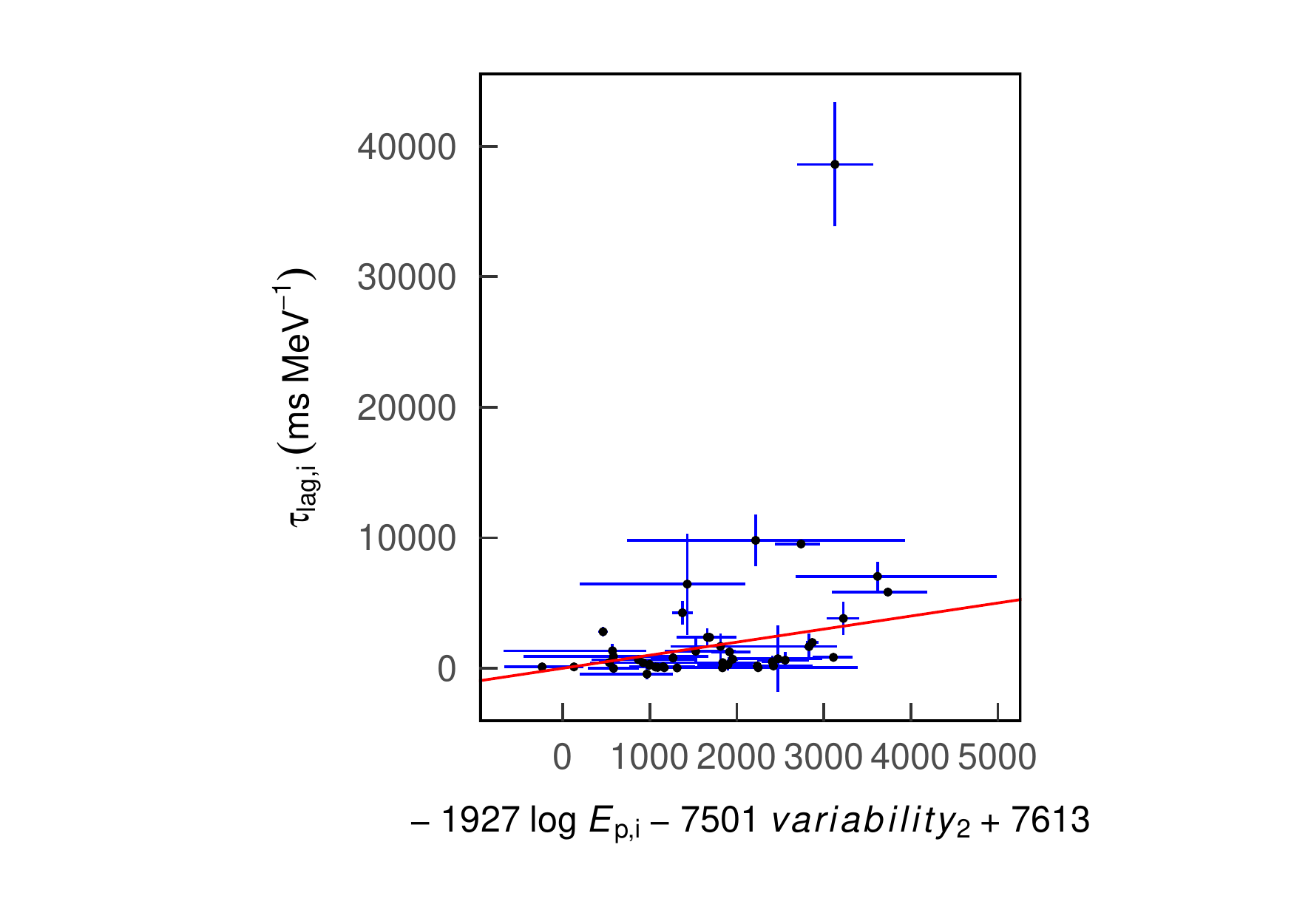}
\caption{Similar to Figure \ref{fig:beta_bandT50ispectral_lagi}, with the three parameters being $\tau_{\rm lag,i}$, $E_{\rm p,i}$ and $variability_{\rm 2}$. The outlier is GRB 980425B.  The linear regression result is $\tau_{\rm lag,i} = (-1927^{\rm +490}_{\rm -500}) \times \log E_{\rm p,i} - (7501^{\rm +3600}_{\rm -3200}) \times variability_{\rm 2} + (7613^{\rm +570}_{\rm -1400})$. The adjusted $R^{2}$ is 0.16.}
\label{fig:variability2Epispectral_lagi}
\end{figure}

The outlier analysis can be apparently shown in figures. We list the results with remarkable linear regression results in Figures \ref{fig:beta_bandT50ispectral_lagi}-\ref{fig:variability2Epispectral_lagi}.

Figures \ref{fig:beta_bandT50ispectral_lagi} and  \ref{fig:beta_bandT90ispectral_lagi} show the outliers in the diagram of  $\tau_{\rm lag,i}$, $T_{\rm 90,i}$ (and $T_{\rm 50,i}$) and {$\beta_{\rm Band}$}, respectively. Both $T_{\rm 90,i}$ and $T_{\rm 50,i}$ indicate the duration of the GRBs. Therefore, these two {analyses} have similar behavior. GRBs 980425B and 030528A are the two outliers, which can be clearly seen from these two figures. The spectral time lag for GRB 030528A is 12.5 $\pm$ 0.5 $\rm s$ between 100-300 $\rm keV$ and 25-50 $\rm keV$ \citep{Schaefer2007}. The spectral time lag for GRB 980425B is 1.46 $\pm$ 0.18 $\rm s$ between 50-100 $\rm keV$ and 25-50 $\rm keV$ \citep{Zhang2009}. Because the energy band is different, we divide the spectral time lag by the difference of two energy band central {values}. $\tau_{\rm lag,i}$ for these two GRBs are 43215 $\pm$ 1729 $\rm ms ~ MeV^{\rm -1}$ and 38605 $\pm$ 4760 $\rm ms ~ MeV^{\rm -1}$, respectively, which are much larger than other GRBs. The detailed data for all the samples can be seen in \citep{Wang2020}. Notice GRB 980425B is the familiar burst which is generally called GRB 980425 \citep[see][for the explanation]{Wang2020}. From the right panels of these two figures, one can see the two outliers have extraordinarily large lags, which should be the reason why they are classified as outliers. From the linear regression results, one can see the spectral time lag is proportional to duration of GRBs, and anti-correlated to the spectral index $\beta$.

Without the two outliers, one finds $\tau_{\rm lag,i} = (-1091^{\rm +152}_{\rm -164}) \times \log L_{\rm pk,52} + (684^{\rm +118}_{\rm -129}) \times \log E_{\rm iso,52} + (703^{\rm +37}_{\rm -85})$. Considering $L_{\rm pk,52} \sim E_{\rm iso,52}/T_{90}$, this relation is actually showing the lag is anti-correlated to the total energy, i.e., with higher total energy, the lag is shorter.

The indication from Figure \ref{fig:E_isoL_pkspectral_lagi} is identified in Figure \ref{fig:E_isoT90ispectral_lagi}. It directly shows the proportional correlation to $T_{90}$ and the anti-correlation to $E_{\rm iso,52}$.

Figure \ref{fig:L_pkalpha_bandspectral_lagi} introduces a new parameter, the spectral index $\alpha$. The lag is positively correlated to $\alpha$, i.e., with bigger $\alpha$, the lag is larger. Remembering the anti-correlation to $\beta$ as shown in Figures \ref{fig:beta_bandT50ispectral_lagi} and  \ref{fig:beta_bandT90ispectral_lagi}, i.e., with smaller $\beta$, the lag is larger. Notice that in general, $\alpha > 0$, and $\beta < 0$, that means the sharper the slope of the spectrum, the larger the lag. This {is confirmed} in Figure \ref{fig:L_pkbeta_bandspectral_lagi}, {where} again $\tau_{\rm lag, i}$ is anti-correlated to $\beta$.

Figure \ref{fig:L_pkE_P_bandispectral_lagi} shows a {positive correlation between the lag and the peak} energy of GRBs. Another new correlation is shown in Figure \ref{fig:variability2Epispectral_lagi}, which is the anti-correlation between the lag and the variability. Because GRB 030528A does not have the value of $variability_{\rm 2}$, we just have one outlier as shown in Figure \ref{fig:variability2Epispectral_lagi}. Both of them have larger adjusted $R^{2}$, indicating the correlation is more reliable.

\subsection{Remarkable outliers without significant linear regression}
\label{subsec:outliers}

\begin{figure}
\centering
\includegraphics[width=0.45\textwidth]{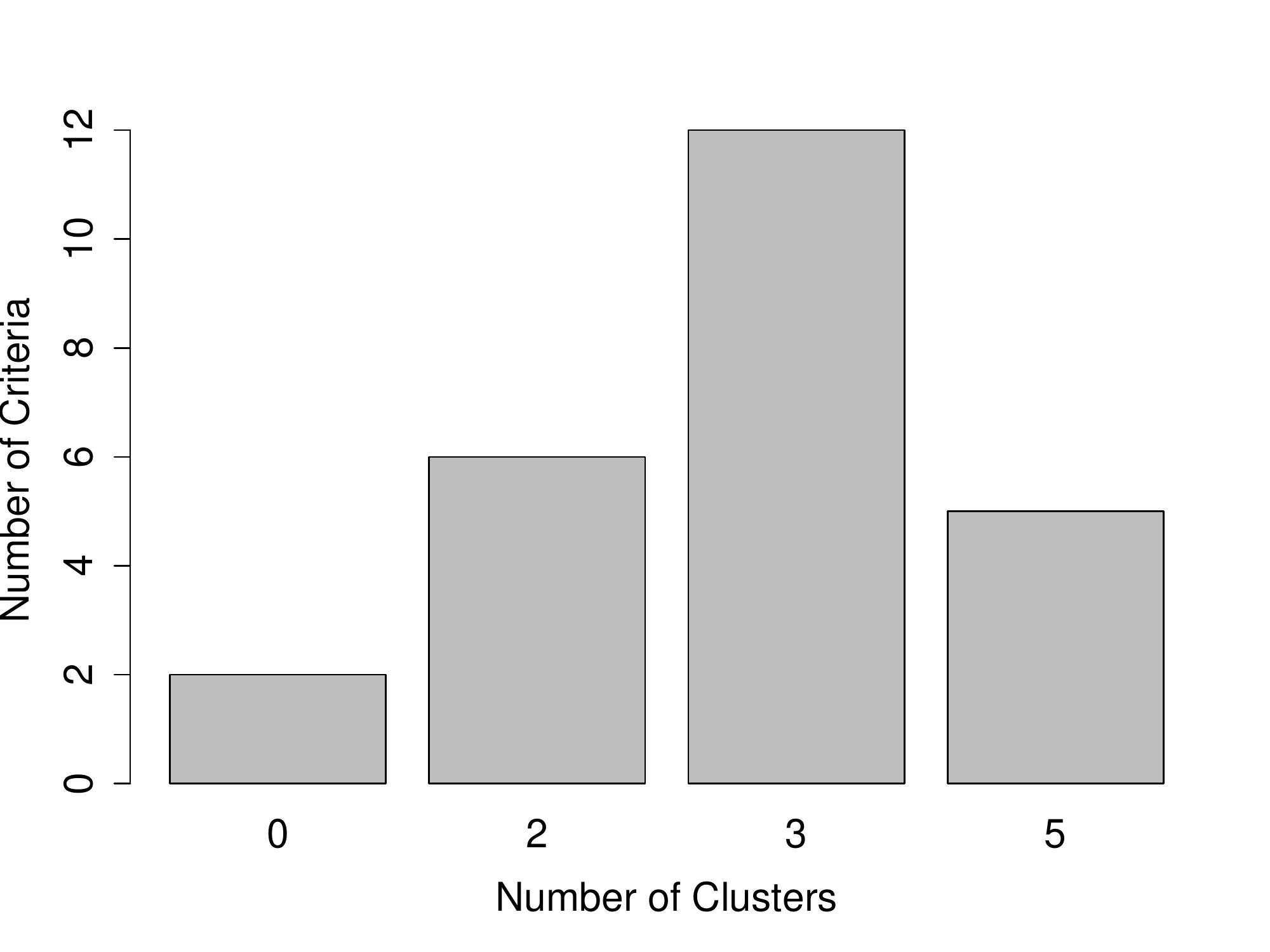}
\includegraphics[width=0.45\textwidth]{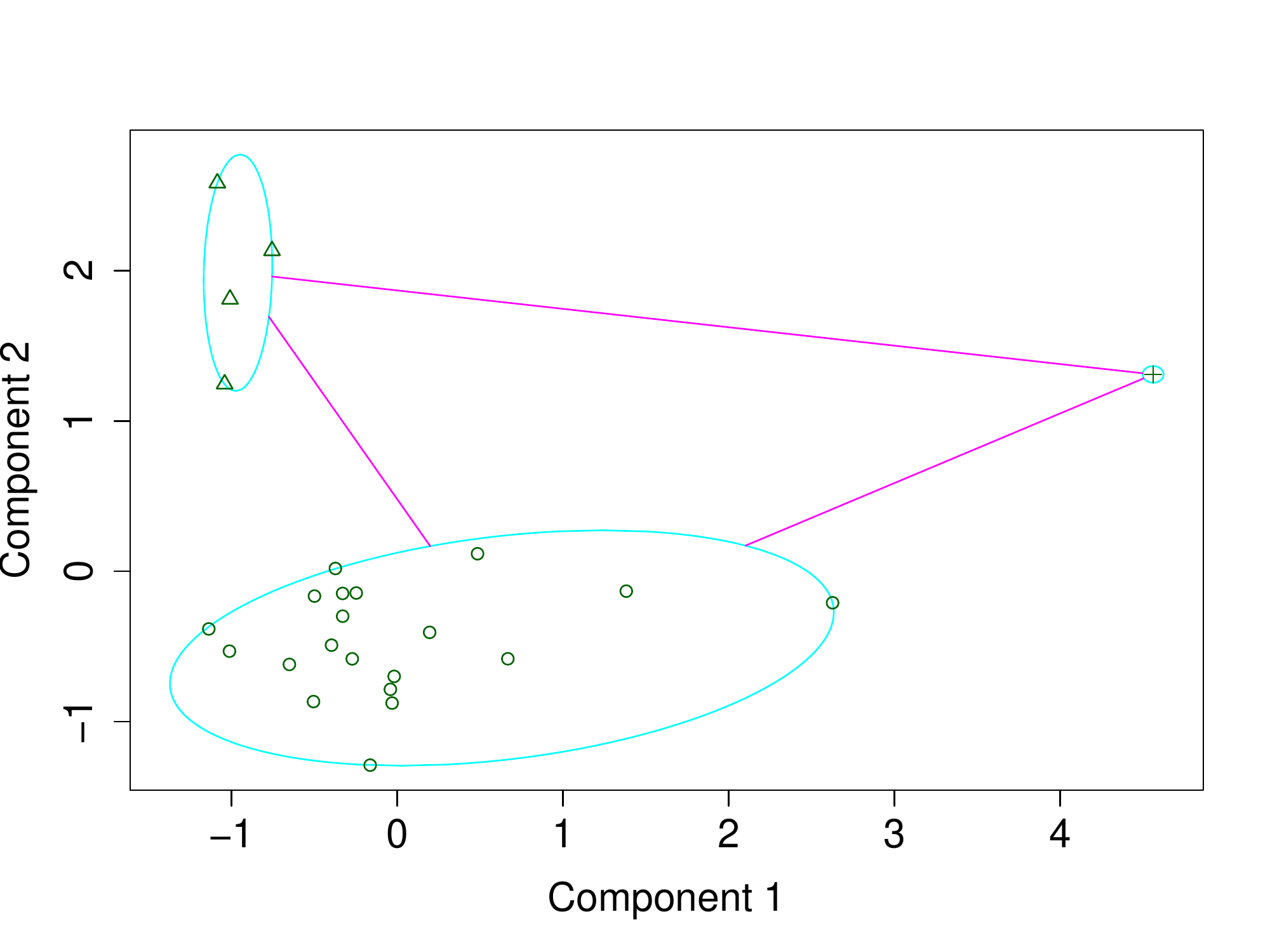}
\caption{PAM result for the clustering analysis for the three parameters being $\tau_{\rm lag,i}$, $\log t_{\rm burst,i}$,  {$\beta_{\rm Band}$}. The histogram is the result of function NbClust, indicating the best cluster number is 3. The right panel shows the PAM result for the clustering analysis. The introduction of every parameter is in Section \ref{sec:method}. The one outlier is 080319B. The four outliers are GRBs 080721A, 090926A, 091127A and 130408A.}
\label{fig:beta_bandspectral_lagilog_t_bursti}
\end{figure}

\begin{figure}
\centering
\includegraphics[width=0.45\textwidth]{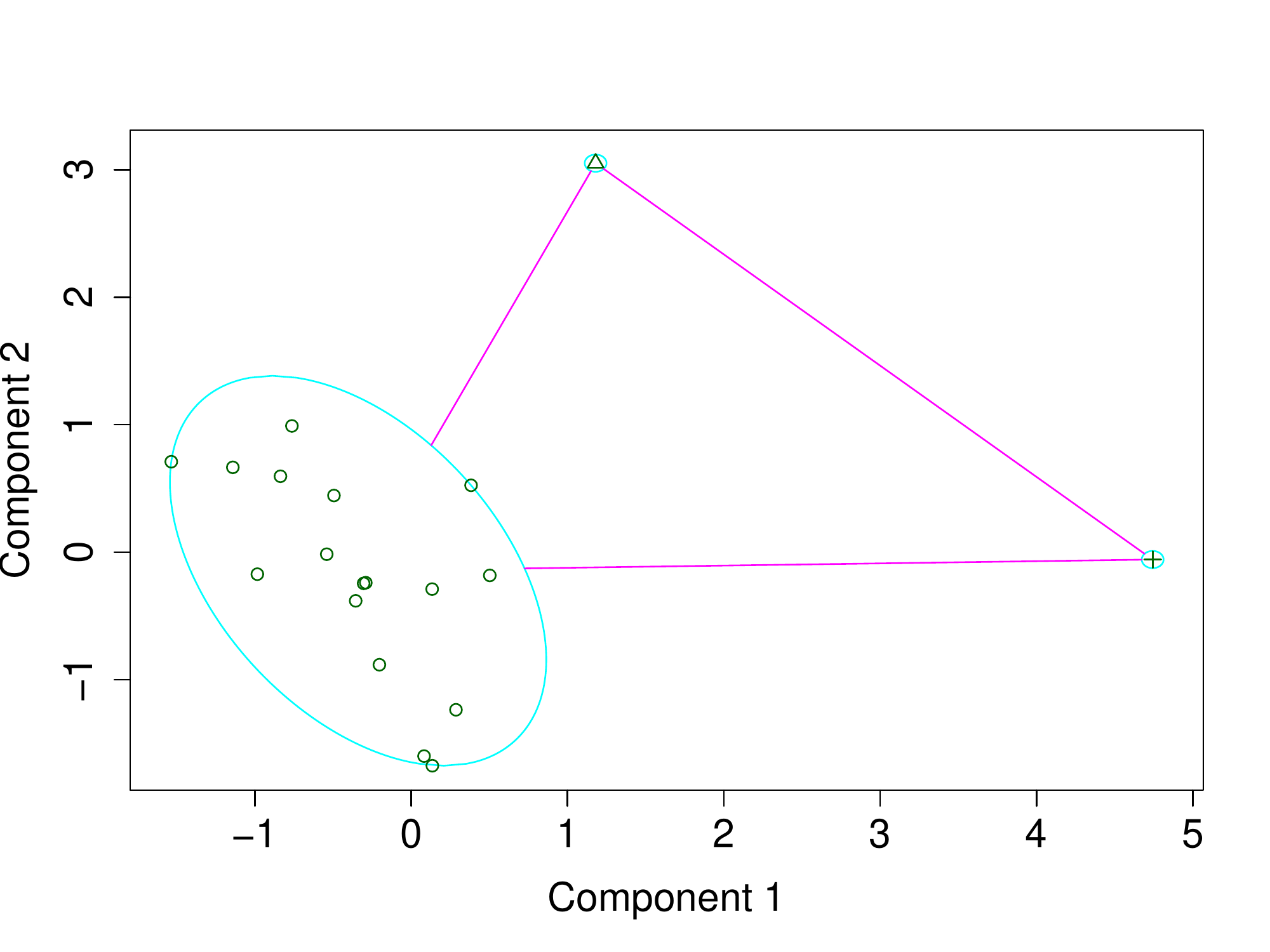}
\includegraphics[width=0.45\textwidth]{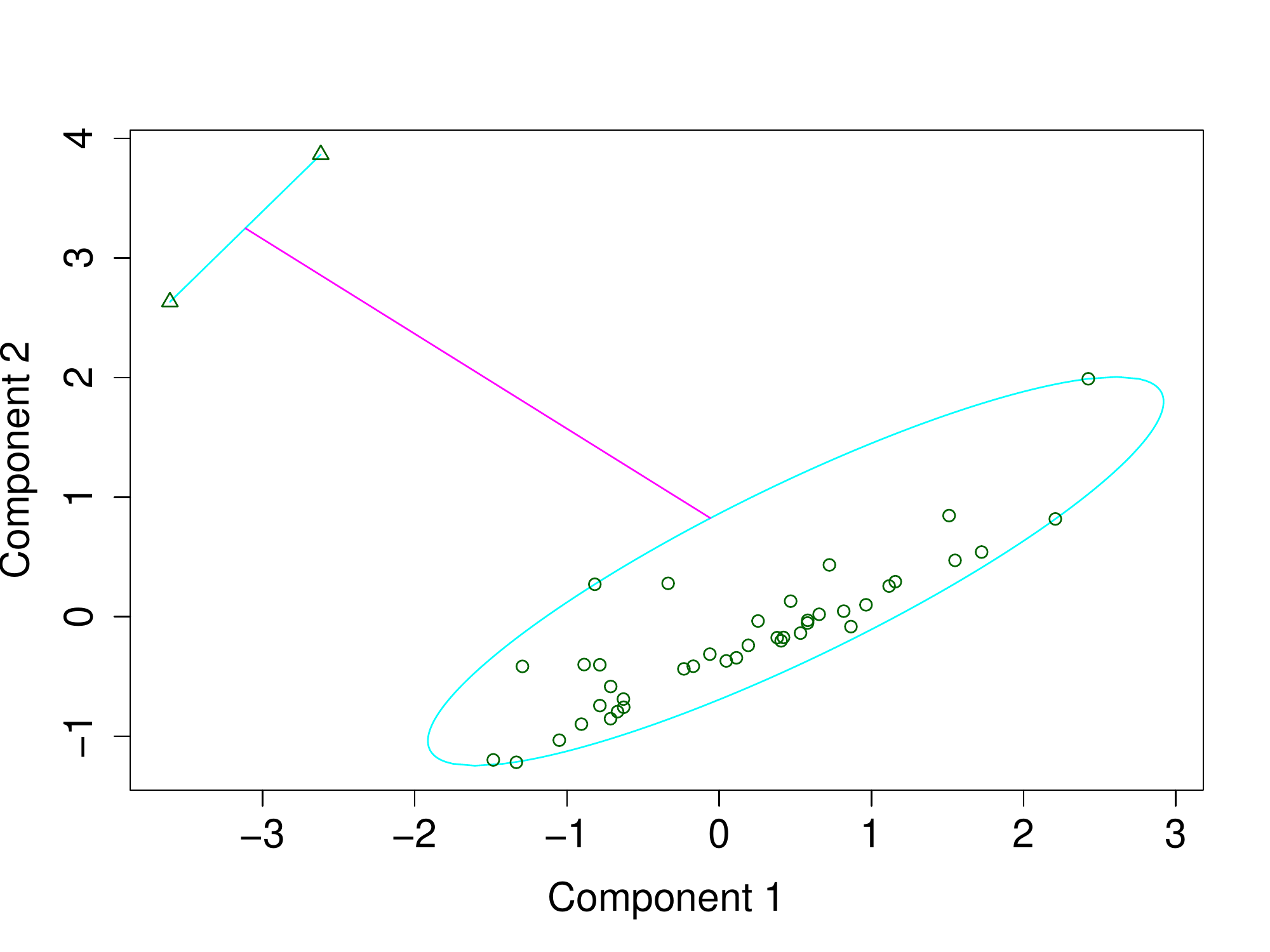}
\caption{Similar to the right panel of Figure \ref{fig:beta_bandspectral_lagilog_t_bursti}. The left panel shows PAM clustering with the three parameters being $\tau_{\rm lag,i}$, Mag, {$\beta_{\rm Band}$}, while the outliers are GRBs 080319B and 081221A. The right panel  shows PAM clustering with the three parameters being $\tau_{\rm lag,i}$, Age, $A_{\rm V}$, while the outliers are GRB 980425B and 030528A.}
\label{fig:beta_bandMagspectral_lagi}
\end{figure}

\begin{figure}
\centering
\includegraphics[width=0.45\textwidth]{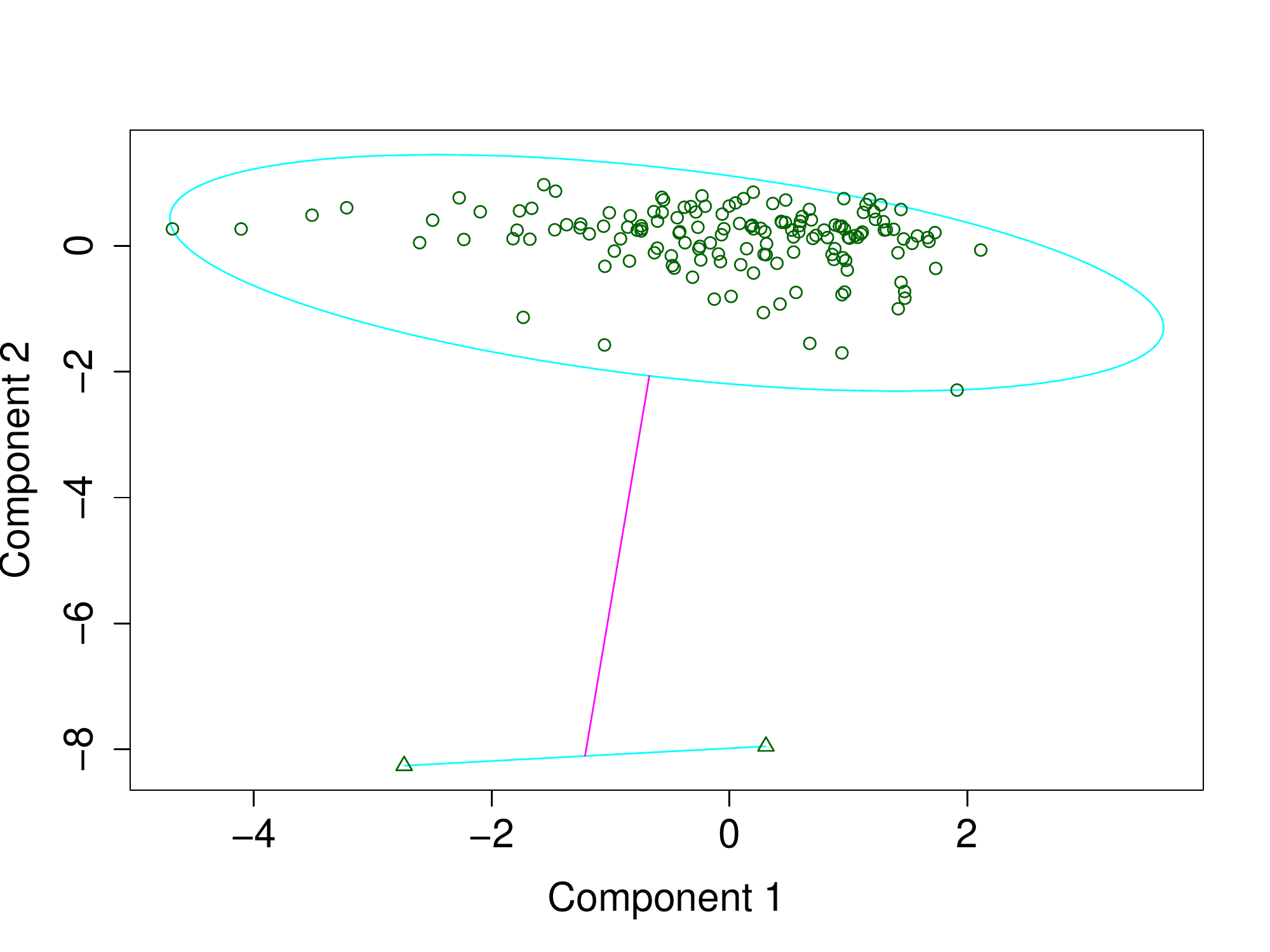}
\includegraphics[width=0.45\textwidth]{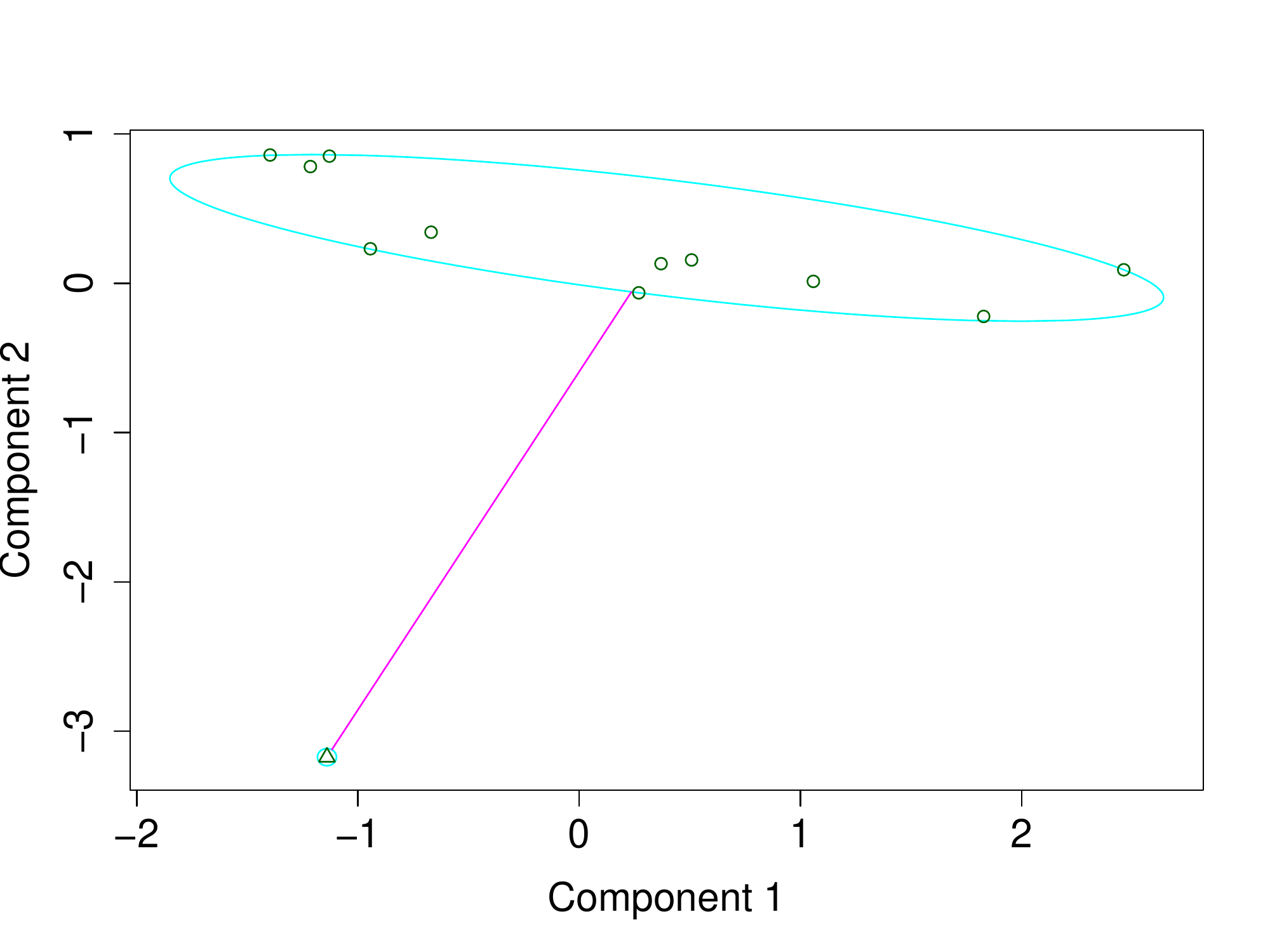}
\caption{Same as Figure \ref{fig:beta_bandMagspectral_lagi}.  The left panel shows PAM clustering with the three parameters being $\tau_{\rm lag,i}$, $T_{\rm R45,i}$, $E_{\rm iso,52}$, while the outliers are GRB 980425B and 030528A. The right panel shows PAM clustering with the three parameters being $\tau_{\rm lag,i}$, $\log t_{\rm burst,i}$, Age, while the outlier is GRB 100621A.}
\label{fig:E_isoT_R45ispectral_lagi}
\end{figure}

\begin{figure}
\centering
\includegraphics[width=0.45\textwidth]{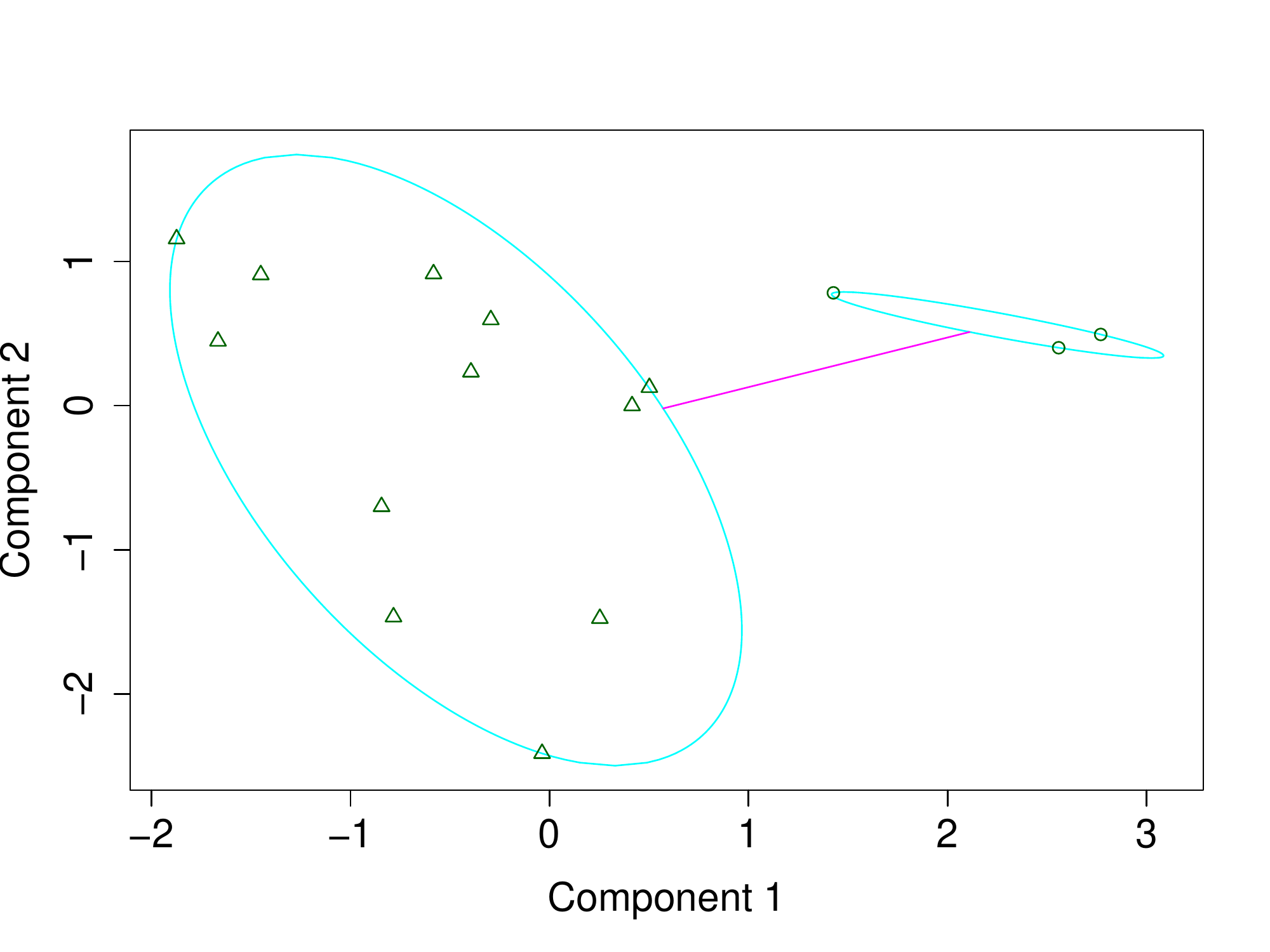}
\includegraphics[width=0.45\textwidth]{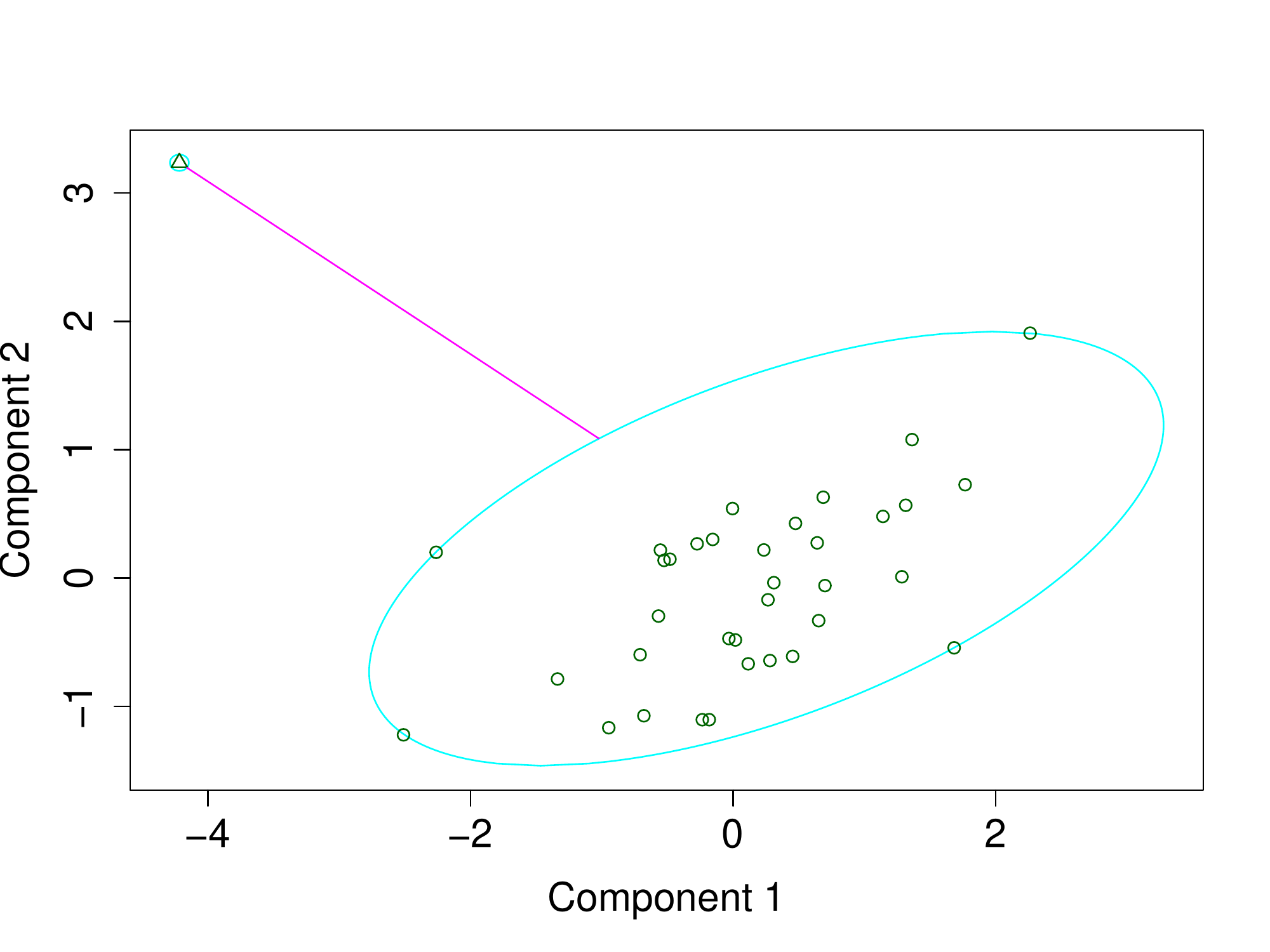}
\caption{Same as Figure \ref{fig:beta_bandMagspectral_lagi}.  The left panel shows PAM clustering with the three parameters being $\tau_{\rm lag,i}$, $t_{\rm radio,pk,i}$, Age, while the outliers are GRBs 010921A, 031203A, 050416A. The right panel shows PAM clustering with the three parameters being $\tau_{\rm lag,i}$, $N_{\rm H}$,  {$\alpha_{\rm Band}$}, while the outlier is GRB 080319B.}
\label{fig:Agespectral_lagit_radio_pki}
\end{figure}

\begin{figure}
\centering
\includegraphics[width=0.45\textwidth]{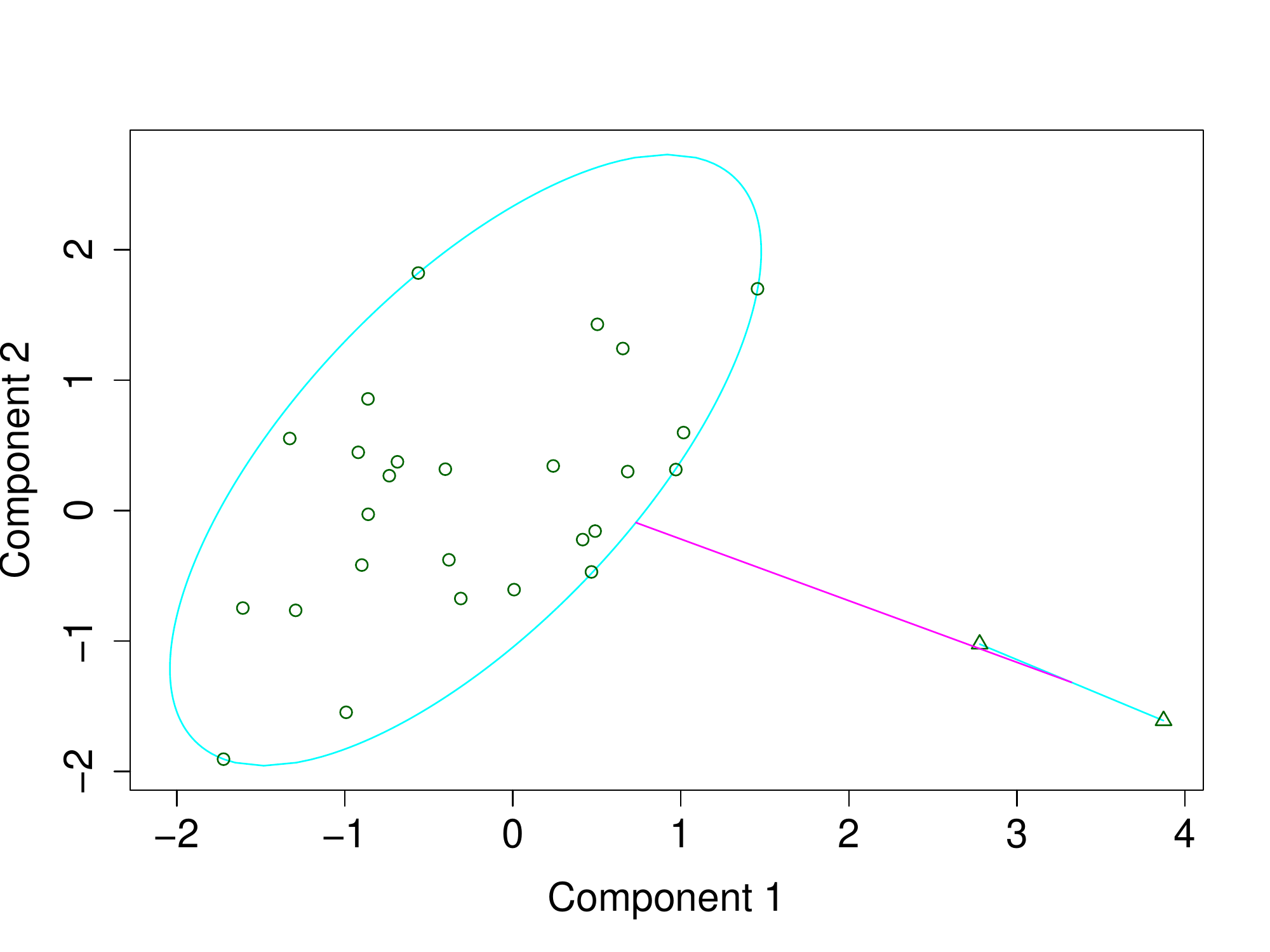}
\includegraphics[width=0.45\textwidth]{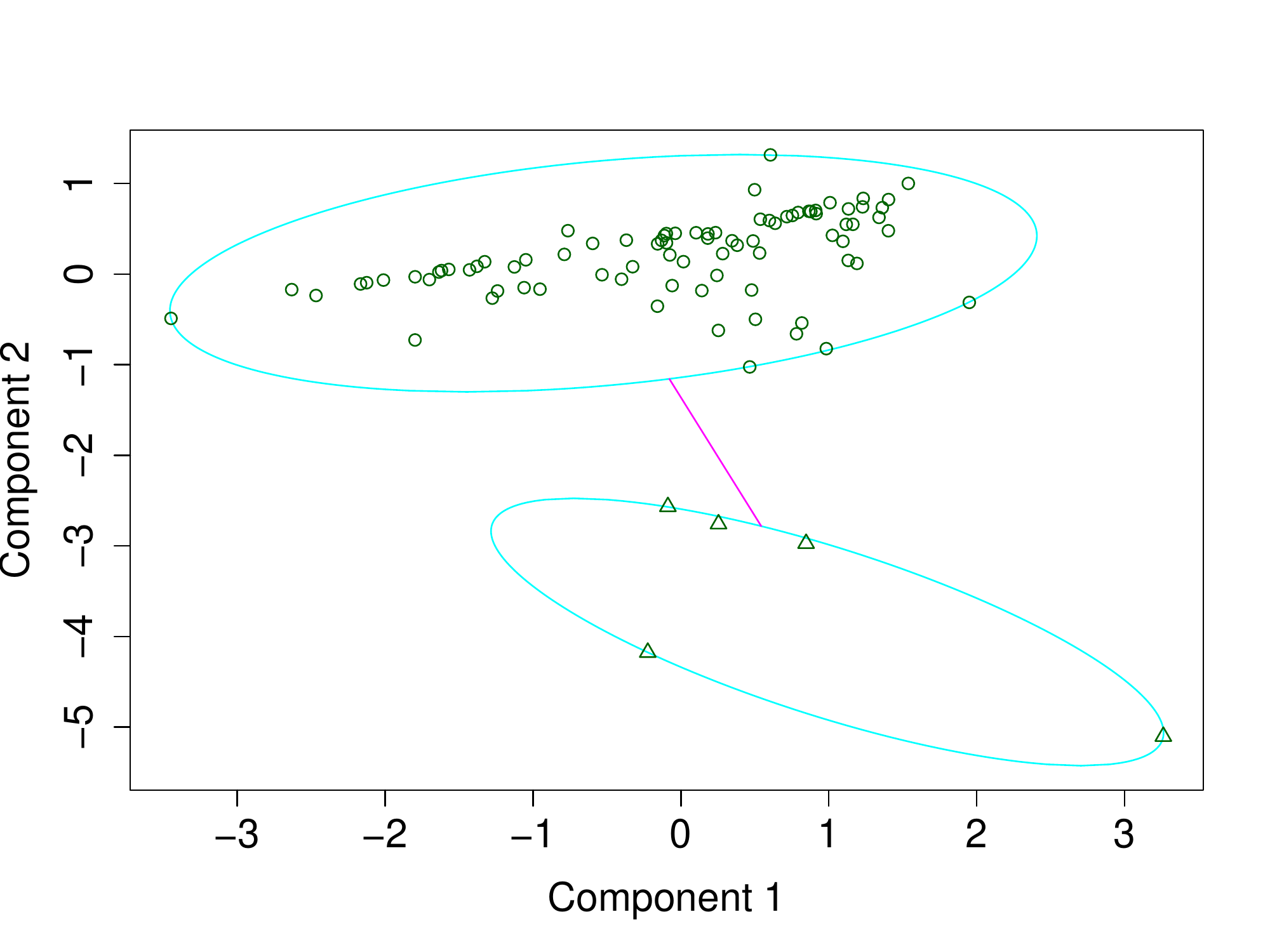}
\caption{Same as Figure \ref{fig:beta_bandMagspectral_lagi}.  The left panel shows PAM clustering with the three parameters being $\tau_{\rm lag,i}$, host galaxy offset,  $\alpha_{\rm cpl}$, while the outliers are GRBs 031203A and 060502A. The right panel shows PAM clustering with the three parameters being $\tau_{\rm lag,i}$, $\beta_{\rm X11hr}$,  $E_{\rm iso,52}$, while the outliers are GRBs 050416A, 060604A, 060605A, 080319B, 100621A.}
\label{fig:alpha_cploffsetspectral_lagi}
\end{figure}

\begin{figure}
\centering
\includegraphics[width=0.45\textwidth]{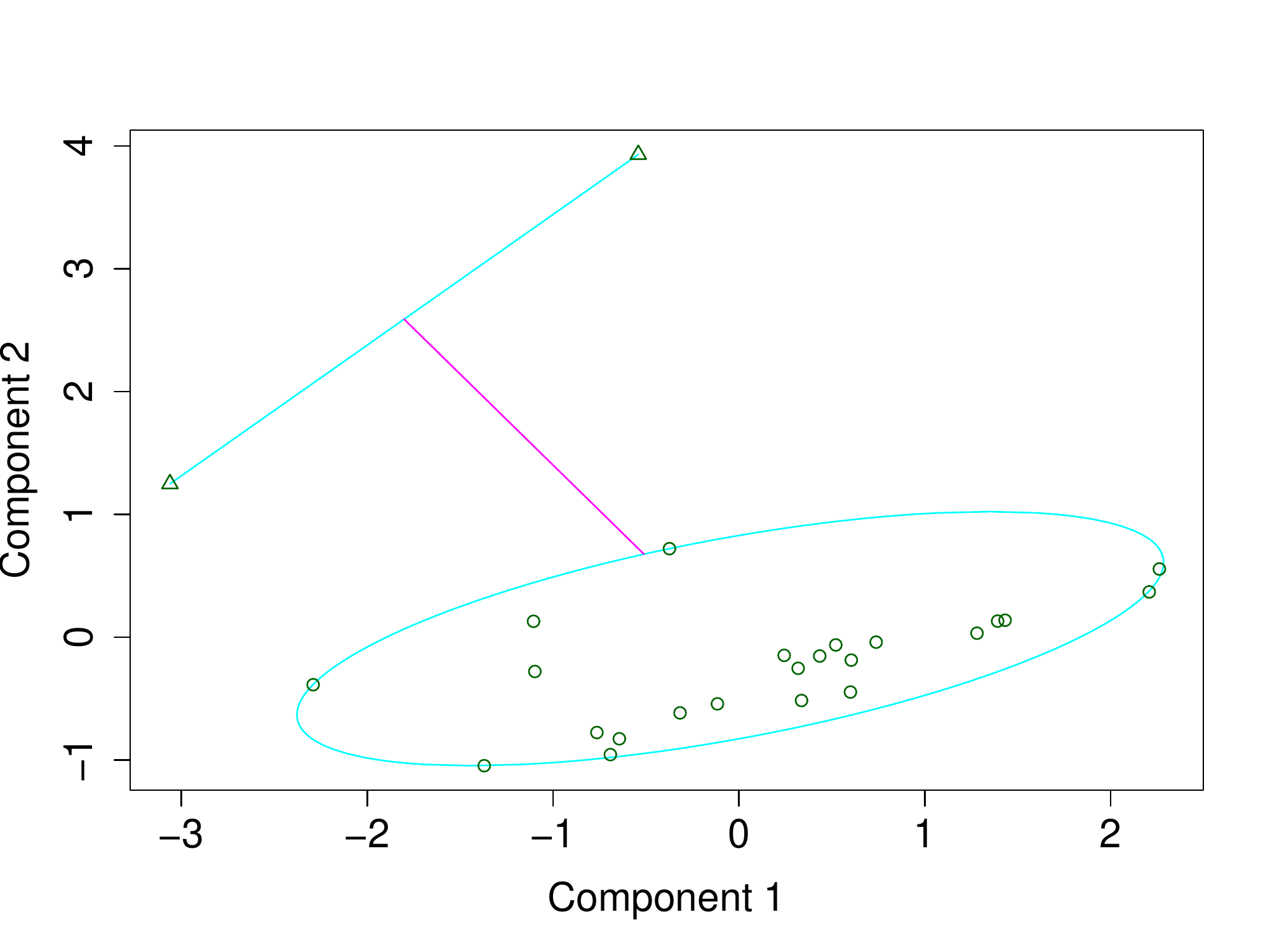}
\includegraphics[width=0.45\textwidth]{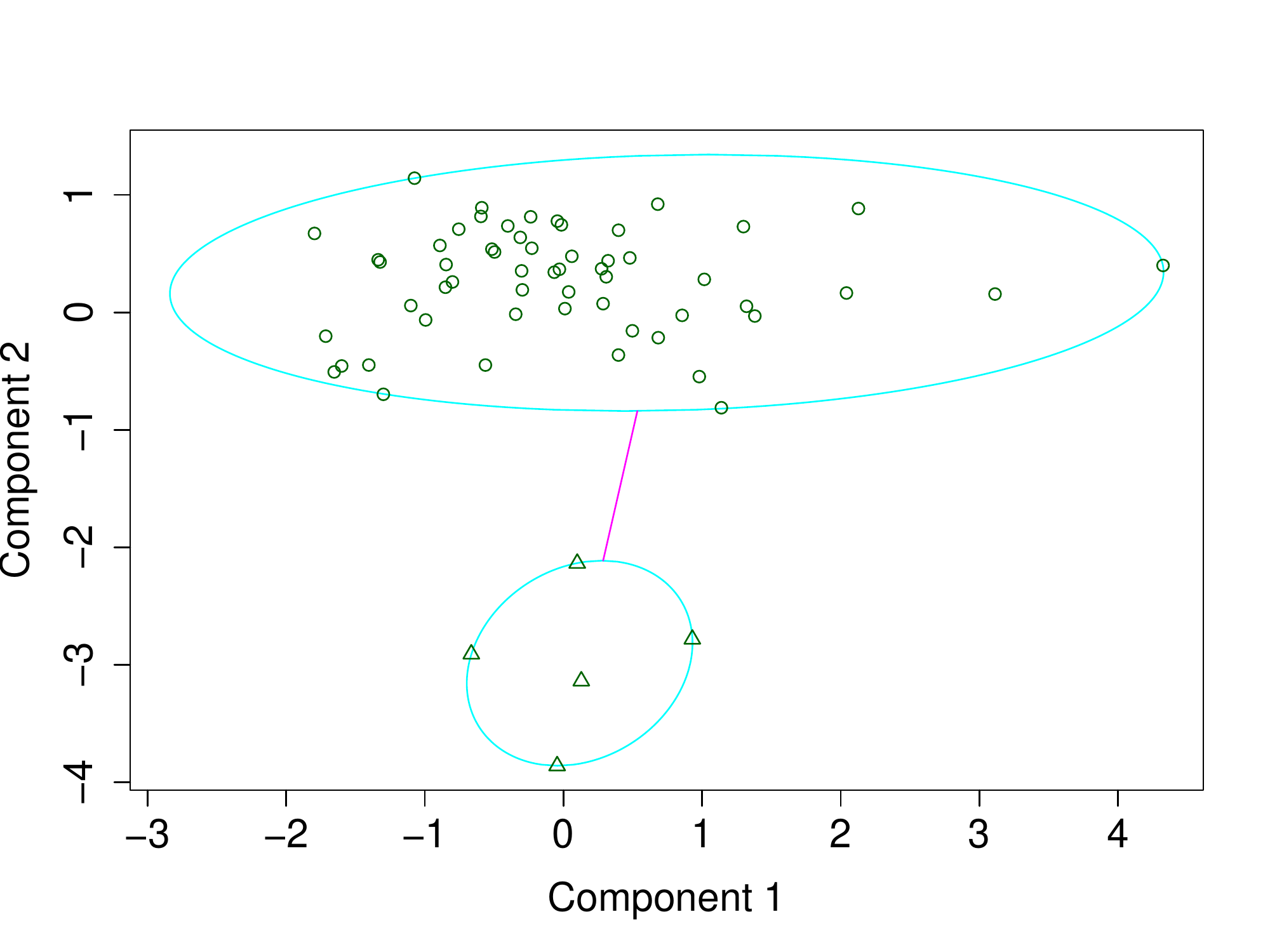}
\caption{Same as Figure \ref{fig:beta_bandMagspectral_lagi}.  The left panel shows PAM clustering with the three parameters being $\tau_{\rm lag,i}$, $\log SSFR$,  host galaxy offset. The outliers are GRBs 031203A, 060505A. The right panel shows PAM clustering with the three parameters being $\tau_{\rm lag,i}$, $T_{\rm 50,i}$,  $\log t_{\rm burst}$. The outliers are GRBs 060607A, 080721A, 090926A, 091127A, 130408A.}
\label{fig:offsetlog_SSFRspectral_lagi}
\end{figure}

Figures \ref{fig:beta_bandspectral_lagilog_t_bursti} to \ref{fig:offsetlog_SSFRspectral_lagi} show the remarkable outliers without significant linear regression {(with p-value greater than 0.05)}, which are clearly obvious. In some of these figures (e.g., right panel of Figure \ref{fig:beta_bandMagspectral_lagi}, and Figure \ref{fig:E_isoT_R45ispectral_lagi}), one can see the sample without outliers shows apparent linear correlations. However, one cannot deduce an intrinsic correlation of the physical parameters, as PAM method only identifies the outliers into apparent large distances.

From these figures, we found more outliers for different combinations of three parameters. The outliers are GRBs 980425B, 010921A, 030528A, 031203A, 050416A, 060502A, 060604A, 060605A, 060607A, 080319B, 080721A, 081221A, 090926A, 091127A, 100621A, and 130408A respectively in different cases.

\begin{figure}
\includegraphics[width=0.45\textwidth]{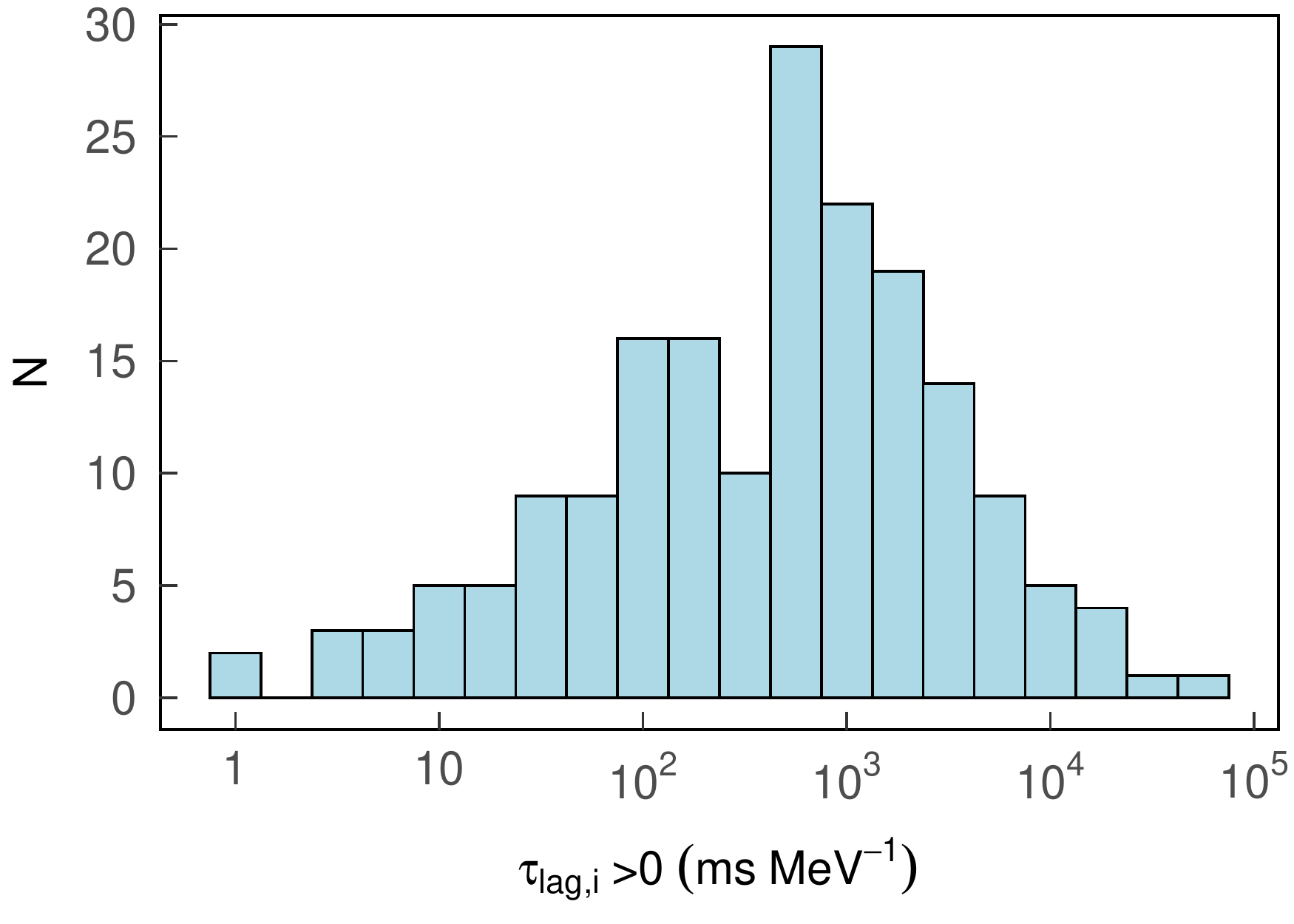}
\includegraphics[width=0.45\textwidth]{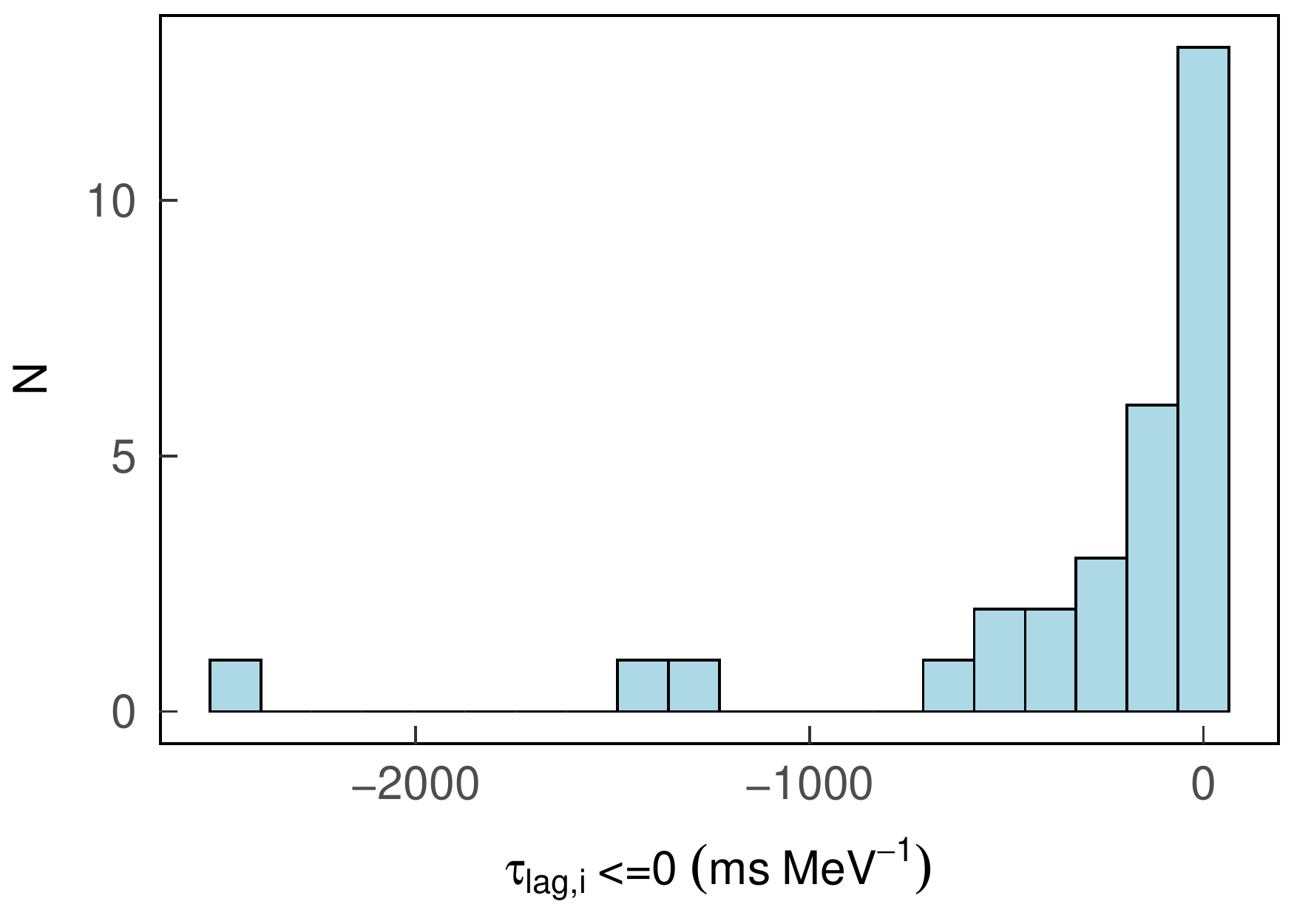}
\caption{The distribution of rest frame spectral time lag $\tau_{\rm lag,i}$. The left panel shows the GRBs with positive values in logarithmic scale. The right panel shows the ones with negative values in normal scale.}
\label{fig:distribution}
\end{figure}

We also show the distribution of $\tau_{\rm lag,i}$ in Figure \ref{fig:distribution}, the outliers are not obvious. Though the quantities are clearly larger for these GRBs, from the distribution itself, they could be inside a single distribution function. Only when they are combined with other parameters, the outliers become obvious. To {show} the distribution more obviously, we choose the logarithmic scale. However, to avoid the negative numbers, we put them in an independent panel as shown in  Figure \ref{fig:distribution}. We also did not consider the error bars in the distribution, as that may rise the positive-negative value problem.

\section{Discussion and conclusions} \label{sec:discuss}
By performing PAM method onto the GRB data, we expect to find some outliers. By removing the outliers, we then expect to find inner correlations with the remaining sample. In this work, we chose the spectral time lag selected sample, together with other parameters, {and we found that, in most cases the outliers are GRBs 980425B and 030528A.} This is mainly because  {they are the two GRBs with} the {largest} spectral lags. However, there are also other outliers in other combinations. That means the value of the lag is not the only {criterion}. In some combinations of other two parameters, there are weak correlations by removing the outliers, while there are also others having no correlations. As there {are} no strong correlations, we can not get very conclusive results from those correlations yet. We expect more combinations can reveal some underline correlations. Once we find tighter correlations, some of those correlations could be used as standard candle  {relations} and could be used for more reliable  {cosmological studies}.

It is interesting that the outliers are always GRBs 980425B and 030528A for those combinations having correlations (see Figures \ref{fig:beta_bandT50ispectral_lagi}-\ref{fig:variability2Epispectral_lagi}), while the others having more outliers. The reason is not clear yet. {GRBs 980425B and 030528A both have very large $\tau_{\rm lag,i}$. In Figures \ref{fig:beta_bandspectral_lagilog_t_bursti}-\ref{fig:offsetlog_SSFRspectral_lagi}, most outliers are not GRBs 980425B and 030528A. It is due to the reason that not all the combinations include GRBs 980425B and 030528A. On the other hand, GRB 980425B is a low-luminosity burst \citep{Rudolph2022}, and GRB 030528A is an X-ray rich burst \citep{Rau2005}. They are both special GRBs.}

With  {the} accumulated data, one can perform both the classification and correlation analysis on the data. One can also first classify the GRBs into several subgroups and look for the correlation for each subgroup. These processes may reveal the underlying nature. Outlier analysis is similar to the classification, while to find the outliers, which is the minority in the whole sample. Those outliers may indicate some special feature of the selected GRBs, which may reveal an independent origin (but much smaller samples), or a very different radiation mechanism. More outlier criteria (more parameter combinations) may reveal that in different aspects. {For example, the soft $\gamma-$ray repeaters, may lie in the sample of conventional GRBs from compact binary mergers or the collapse of massive stars, and may be classified as outliers with a certain set  {of} correctly selected parameters.}
If the outliers are real strangers, one may want to remove {them} from the whole sample.
By omitting the outliers, the remaining sample may obey some laws, which can also be used to study the physics of GRBs. With the remaining sample, one may want to do {a} similar the classification and correlation analysis. In the future more comprehensive study, new patterns might be revealed.




\funding{This work is in part supported by the National Natural Science Foundation of China (Grant Nos. U1931203 and 12041306), by the National Key R\&D Program of China (2021YFC2203100). We also acknowledge the science research grants from the China Manned Space Project with No. CMS-CSST-2021-B11.}

\dataavailability{The data used in this work are publicly available in the machine readable table in \cite{Wang2020}.}


\conflictsofinterest{The authors declare no conflict of interest.}



\begin{adjustwidth}{-\extralength}{0cm}

\reftitle{References}


\bibliography{export-bibtex}


%


\end{adjustwidth}
\end{document}